\documentclass[iop]{emulateapj}

\usepackage{gensymb}
\usepackage{natbib}
\usepackage{amsmath}
\usepackage{xcolor}
\usepackage{rotating}

\shorttitle{First Larson core: 3D chemo-dynamical simulation}
\shortauthors{Hincelin et al.}

\begin{document}

\title{Chemical and physical characterization of collapsing low-mass prestellar dense cores}

\author{U. Hincelin}
\affil{Department of Chemistry, University of Virginia, Charlottesville, VA 22904, USA}
\email{ugo.hincelin@gmail.com}

\author{B. Commer\c con}
\affil{Ecole Normale Sup{\'e}rieure de Lyon, CRAL, UMR 5574 du CNRS, Universit{\'e} Lyon I, 46 All{\'e}e d'Italie, 69364 Lyon cedex 07, France}

\author{V. Wakelam, F. Hersant, S. Guilloteau}
\affil{Univ. Bordeaux, LAB, UMR 5804, F-33270, Floirac, France and \\
CNRS, LAB, UMR 5804, F-33270, Floirac, France}

\and

\author{E. Herbst}
\affil{Departments of Chemistry and Astronomy, University of Virginia, Charlottesville, VA 22904, USA}

\begin{abstract}
The first hydrostatic core, also called the first Larson core, is one of the first steps in low-mass star formation, as predicted by theory.
With recent and future high performance telescopes, details of these first phases become accessible, and observations may confirm theory and even bring new challenges for theoreticians.
In this context, we study from a theoretical point of view the chemical and physical evolution of the collapse of prestellar cores until the formation of the first Larson core, in order to better characterize this early phase in the star formation process.
We couple a state-of-the-art hydrodynamical model with full gas-grain chemistry, using different assumptions on the magnetic field strength and orientation.
We extract the different components of each collapsing core (i.e., the central core, the outflow, the disk, the pseudodisk, and the envelope) to highlight their specific physical and chemical characteristics.
Each component often presents a specific physical history, as well as a specific chemical evolution.
From some species, the components can clearly be differentiated.
The different core models can also be chemically differentiated.
Our simulation suggests some chemical species as tracers of the different components of a collapsing prestellar dense core, and as tracers of the magnetic field characteristics of the core.
From this result, we pinpoint promising key chemical species to be observed.
  
\end{abstract}


\section{Introduction}

The collapse of a prestellar dense core, a dense region of a molecular cloud, leads to the formation of the first hydrostatic core (FHSC), also called the first Larson core \citep{larson_numerical_1969}.
According to the theory of magnetised dense core collapse \citep[e.g.,][]{joos_protostellar_2012}, this process can result in the formation of a rotationally supported disk surrounding the first Larson core, and is associated with the beginning of the launch of the outflow.
A pseudodisk, i.e. a non-rotationally supported disk, can also be formed during the process \citep{galli_collapse_1993-1,galli_collapse_1993}.
Depending on the intensity of the magnetic field, fragmentation may occur as well \citep[e.g.,][]{commercon_protostellar_2010}.
An adiabatic contraction continues to enhance the temperature of the core, and once it reaches $\sim2000$~K, the endothermic reaction of H$_2$ dissociation leads to a second collapse.
The second hydrostatic core, or the second Larson core, then forms, which marks the end of the prestellar phase and the beginning of the protostellar phase \citep{andre_prestellar_2000}.
The newly formed protostar will then continue to accrete matter through a disk, until the fusion of deuterium and hydrogen indicate its transition to a star.

The first Larson core has not been detected with certainty, mainly because it has a short lifetime ($\sim 1000$~yr) and is embedded \citep{commercon_synthetic_2012,tomida_radiation_2013}.
However a few candidates have recently been reported \citep{belloche_evolutionary_2006,chen_l1448_2010,chen_submillimeter_2012,enoch_candidate_2010,dunham_detection_2011,pineda_enigmatic_2011,pezzuto_herschel_2012,tsitali_dynamical_2013,huang_probing_2013,hirano_two_2014,gerin_nascent_2015}. 
The increasing resolution of the instruments, specially ALMA\footnote{Atacama Large Millimeter/submillimeter Array}, allows astronomers to see many more details of the early phases of star forming regions than ever before \citep[e.g.,][]{lee_alma_2014,friesen_revealing_2014} and the complete ALMA interferometer will be soon able to probe the FHSC scale in nearby star-forming regions \citep{commercon_synthetic_2012-1}.
Therefore, an effort in computational modeling to couple state-of-the-art physical models with full gas-grain chemical codes is one important step towards a better understanding of the physics, the chemistry, as well as the observations of these regions.

The chemistry of collapsing cores has been studied using zero and one-dimensional models \citep{ceccarelli_far-infrared_1996,rodgers_chemical_2003,doty_physical-chemical_2004,lee_evolution_2004,garrod_formation_2006,aikawa_molecular_2008,garrod_complex_2008}. 
\cite{van_weeren_modeling_2009} and \cite{visser_chemical_2009,visser_chemical_2011} presented studies of two-dimensional chemical evolution during the collapse of a molecular cloud to form a low-mass protostar and its surrounding disk.
As a first step to study the coupling between the dynamics of the formation of the first Larson core and its chemical composition, without assuming any symmetry in the physical structure, \cite{furuya_chemistry_2012} performed three-dimensional radiation-hydrodynamical (3D-RHD) simulations coupled with a full gas-grain chemistry.
\cite{hincelin_survival_2013} (designated hereafter by HW13) performed the same kind of simulation, taking into account the effect of the magnetic field on the dynamics of the collapse (3D-RMHD simulation), and studied the formation of the young rotationally supported disk around the first Larson core.

We present in this paper an extension of the work done by HW13.
We study the physical and chemical evolution of the other components of a collapsing dense core besides the disk, i.e. the pseudodisk, the outflow, the collapsing envelope, and the central core (i.e. the {{{FHSC}}} core itself).
In order to study the impact of a high magnetization level on the dynamics and the chemistry, we also present an extra simulation of the collapse of a highly magnetized dense core.
In Section~\ref{sec:model}, we describe our modeling.
We present our results, a physical and a chemical characterization, in Section~\ref{sec:results}.
In Section~\ref{subsec:tracers}, we discuss the chemical distinction that can be made between the different core models, and the different components within a given core, and identify the most promising chemical tracers.
In Section~\ref{sec:obs}, we do some comparison with recent observations of FHSC candidates.
Finally, we conclude our study in Section~\ref{sec:conclusion}.

\section{Modeling}
\label{sec:model} 

We used the gas-grain chemical code {\ttfamily{NAUTILUS}} \citep{hersant_cold_2009} with the physical structure of a collapsing dense core computed by the adaptive mesh refinement code {\ttfamily{RAMSES}} \citep{teyssier_cosmological_2002}.
The codes and the method used are the same ones as in HW13.

\subsection{The physical code {\ttfamily{RAMSES}}}
\label{subsec:RAMSES} 

The physical structure is derived using the RMHD solver of {\ttfamily{RAMSES}} which integrates the equations of ideal magnetohydrodynamics \citep{fromang_high_2006} and of RHD using the grey flux-limited-diffusion approximation \citep{commercon_radiation_2011}.
The initial conditions are those used in \cite{commercon_synthetic_2012}: a 1~M$_\odot$ sphere of uniform density $\rho_0=3.97\times 10 ^{-18}$~g~cm$^{-3}$ -- corresponding to a molecular hydrogen number density $n_{\rm H_2}\simeq~10^6$~cm$^{-3}$ -- and a temperature $T_0=11$~K.
In order to initiate fragmentation, an azimuthal perturbation is introduced in the initial density following 
\begin{equation}
\rho = \rho_0 \left( 1+A\cos \left( m \theta \right) \right),
\end{equation}
with an amplitude $A=10\%$, and a perturbation $m=2$.
$\theta$ is the azimuthal angle (in cylindrical coordinates).
{The sphere (radius $r_0\simeq 3300$~AU) is in rigid body rotation about the z-axis, and threaded by a uniform magnetic field parallel to the rotation axis.}
The initial rotational rate $\Omega_0$ is expressed in terms of $\beta$, the rotational energy to gravitational energy ratio
\begin{equation}
\beta = \frac{\frac{1}{5} M_0 r_0^2 \Omega_0^2}{\frac{3}{5}\frac{G M_0^2}{r_0}} = 0.045,
\end{equation}
where $G$ is the gravitational constant and 
$M_0$ is {{{the initial mass of 1~M$_\odot$}}}.
The strength of the magnetic field $\mu$ is expressed in terms of the mass-to-flux to critical mass-to-flux ratio
\begin{equation}
\mu=\frac{\frac{M_0}{\Phi}}{\left(\frac{M_0}{\Phi}\right)_\mathrm{c}},
\end{equation}
{{{where}}} $\Phi$ is the magnetic flux threading the cloud.
The critical mass-to-flux ratio of a cloud with a uniform density and magnetic field is given by
\begin{equation}
\left(\frac{M_0}{\Phi}\right)_\mathrm{c}=\frac{c_1}{3\pi}\sqrt{\frac{5}{G}},
\end{equation}
where $c_1=0.53$ \citep{mouschovias_note_1976}.
In this study, we extended the magnetization level to a third value compared to HW13, so that there are a low one ($\mu=200$, very close to a pure hydrodynamical case), a moderate one ($\mu=10$), and a high one ($\mu=2$), that correspond to an initial magnetic field equal to $\sim 3$, $\sim 53$, and $\sim 265$~$\mu$G respectively\footnote{The initial magnetic flux $\Phi_0$ can be linked to the initial magnetic field $B_0$ by $\Phi_0=\pi B_0 r_0^2$.}.
We choose the values of $\mu$ to be consistent with the observations \citep[e.g.][]{heiles_magnetic_2005,falgarone_cn_2008,li_link_2014} and to reproduce the variety of components and dynamics that can be found in collapsing dense cores.
For $\mu=10$, we use two different initial configurations: an angle $\Theta=0\degree$ between magnetic field lines and the rotation axis of the sphere, and another angle $\Theta=45\degree$.
{The non-aligned rotation and magnetic field allows an easier formation of the disk \citep{joos_protostellar_2012}.}
These values are used in total four different models: MU10$\Theta$0, MU10$\Theta$45, MU200$\Theta$0, and MU2$\Theta$0 (see Table~\ref{tab:model3D_liste_modeles}).

\begin{table}[thb]
\begin{center}
\caption{{Magnetic fields strength and orientations used for the current simulations.} \label{tab:model3D_liste_modeles}}
\begin{tabular}{lll}
\tableline
\tableline
Model & $\mu$ & $\Theta$ \\
\tableline
MU10$\Theta$0  & 10 & 0 \\
MU10$\Theta$45 & 10 & 45 \\
MU200$\Theta$0 & 200 & 0 \\
MU2$\Theta$0 & 2 & 0 \\
\tableline
\end{tabular}
\end{center}
\tablecomments{$\Theta$ is given in \degree. The lower $\mu$, the higher the magnetic field strength.}
\end{table}

\begin{figure*}
\centering
\includegraphics[width=1.0\linewidth]{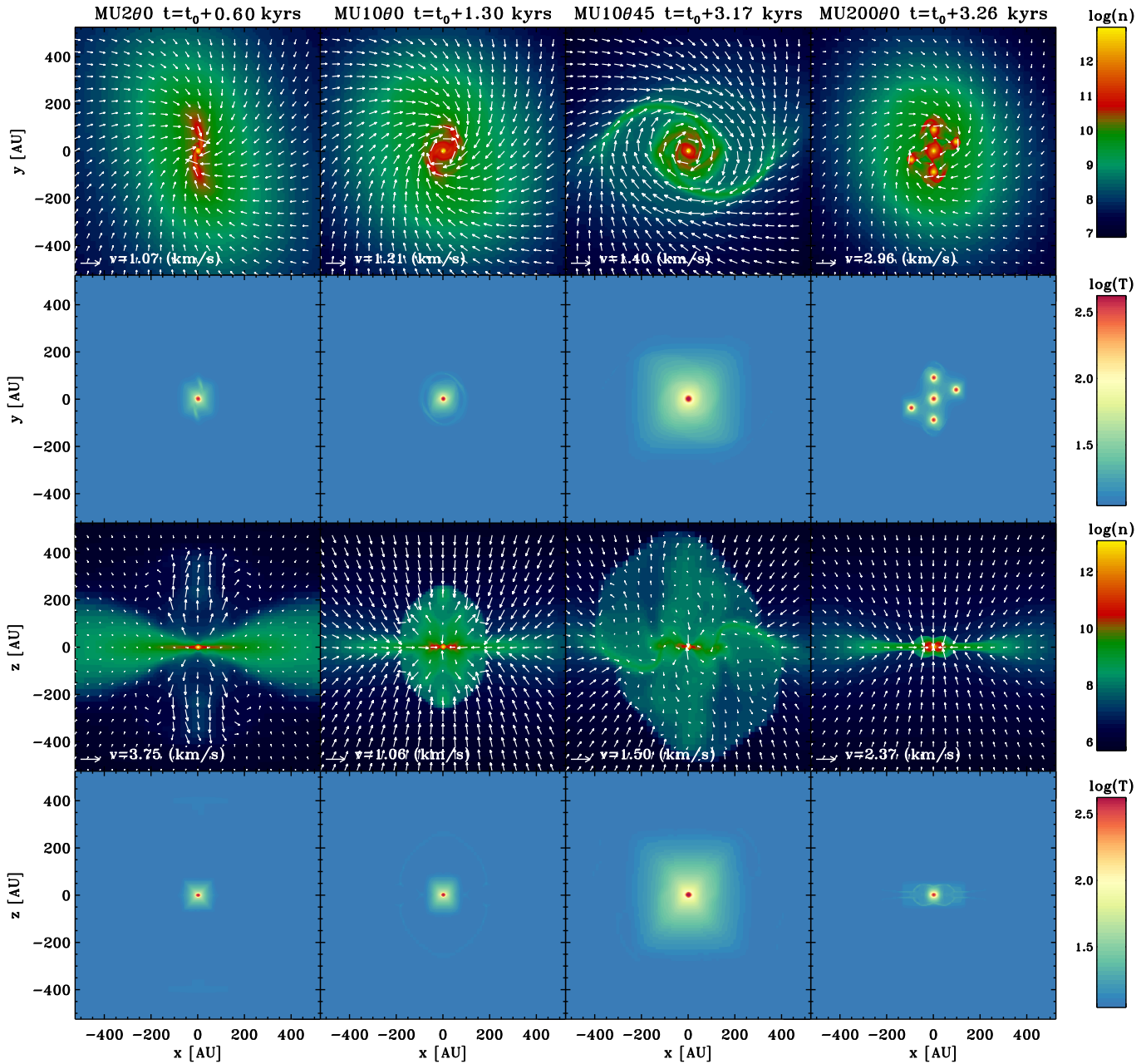}
\caption{{Temperature and density maps computed by {\ttfamily{RAMSES}}, at the end of our simulations, for MU2$\Theta$0, MU10$\Theta$0, MU10$\Theta$45, and MU200$\Theta$0 models.
Information is displayed in the plane $z=0$~AU (top two rows) and in the plane $y=0$~AU (bottom two rows).
Density maps also show velocity field.
The projected vector length indicates velocity value.
See text for the definition of $t_{\rm 0}$}.}
\label{fig:ramses_output}
\end{figure*}

We computed the evolution of the collapsing dense core throughout the first stage of collapse and first Larson core lifetime \citep{larson_numerical_1969}, {which is $\sim 4\times 10^4$~yr} for these four models.
Figure~\ref{fig:ramses_output} shows temperature and density maps of the collapsing core at the end of the simulations.
{The maximum level of refinement is reached at $t_{\rm 0}$, which roughly corresponds to the formation of the FHSC \citep{commercon_synthetic_2012}.
See Table~\ref{tab:times_simulation} for numerical values.}

Our initial conditions allow us to reproduce the diverse components of a collapsing dense core: a central core with or without fragmentation, a disk, a pseudodisk, an outflow, and an envelope (see Section~\ref{subsubsec:components_charac}).
Starting from a Bonnor-Ebert like density profile also reproduces these components.
The initial conditions will mainly impact the details of the component formation  (see \cite{machida_conditions_2014} for a detailed study using a non-ideal MHD simulation).

\subsubsection{Ideal versus non-ideal MHD}
The physical model {\ttfamily{RAMSES}}  integrates the equations of ideal magnetohydrodynamics, assuming the medium to be a perfect
conductor.
The matter is then frozen on the magnetic lines.
If we consider a medium that is not a perfect conductor, that is to say, in the case of non-ideal (or resistive) magnetohydrodynamics, then the neutral matter will be able to cross magnetic field lines.
The ideal MHD approximation has a strong limitation, in particular with regard to the ionization deep inside the collapsing core, which may impact the formation of disks \citep[see e.g.][]{padovani_role_2014}.
The friction between ions and neutrals also generates a net velocity difference between these two families of species that may be sufficient to overcome activation barriers, thereby generating efficient chemical reaction pathways, but this process is still far from being implemented in models.
However, recent work has been done to integrate non-ideal processes in {\ttfamily{RAMSES}}  \citep{masson_incorporating_2012,masson_ambipolar_2015} and future work will be needed to continue to explore the effects on the dynamics and the chemistry.

\begin{table}[thb]
\begin{center}
\caption{Times of simulation (yr).}
\label{tab:times_simulation}
\begin{tabular}{lll}
\tableline
\tableline
Model & $t_{\rm 0}$ & Final time \\
\tableline
MU10$\Theta$0  & 3.57(4) & 3.70(4) \\
MU10$\Theta$45 & 3.60(4) & 3.92(4) \\
MU200$\Theta$0 & 3.54(4) & 3.86(4) \\
MU2$\Theta$0   & 5.04(4) & 5.10(4) \\
\tableline
\end{tabular}
\tablecomments{$\rm a(b)$ means $a\times 10^b$.}
\end{center}
\end{table}

\subsection{The chemical code {\ttfamily{NAUTILUS}}}
\label{subsec:nls} 

{\ttfamily{NAUTILUS}} solves the kinetic equations of gas-phase and grain surface chemistry, and takes into account interactions between both phases (adsorption and thermal and non-thermal desorption).
{The two-phase rate-equation approach is used, in which no distinction is made between the inner and surface layers of the ice mantle.}
The code is based on \cite{hasegawa_new_1993}, is written in Fortran~90, and uses the LSODES solver, part of ODEPACK \citep{hindmarsh_odepack_1983}.
The rate equations follow \cite{hasegawa_models_1992} and \cite{caselli_proposed_1998}.
More details of the physical and chemical processes included in the code are given in a benchmark paper by \cite{semenov_chemistry_2010}.
The chemical network, adapted from \cite{garrod_non-thermal_2007}, follows the chemistry of 458 gas-phase species and 196 species on grains, and includes 6210 reactions (4447 gas-phase reactions and 1763 grain-surface and gas-grain interaction reactions).
The gas-phase network has been updated according to the recommendations from the experts of the KIDA database\footnote{\samepage
KInetic Database for Astrochemistry \citep{wakelam_kinetic_2012}.\\
\url{http://kida.obs.u-bordeaux1.fr}} (current update in October 2011).
{Note that high temperature reactions, for the range 300 to 800~K, were not included at that time, but this limitation may only impact the very center of the FHSC.}
An electronic version of our network is available at \url{http://kida.obs.u-bordeaux1.fr/models} (network from \cite{hincelin_survival_2013}).

To compute the initial chemical composition of the sphere, we ran {\ttfamily{NAUTILUS}} for dense cloud conditions (10~K, total hydrogen density of $2\times 10^5$~cm$^{-3}$, cosmic-ray ionization rate of $1.3\times 10^{-17}$~s$^{-1}$ and a visual extinction of 30) for a time $t\sim 6\times 10^5$~yr.
The chosen time $t$ corresponds to the maximum agreement between observations of molecular clouds and our simulations (see Fig.3 of \cite{hincelin_oxygen_2011}).
At that stage the physical condition is fixed, which means we do not treat the complex formation process of the molecular cloud. Note that the initial density for the chemistry is smaller than the one for the hydro-dynamical simulations. 
The goal of this initial stage is only to have a reliable initial chemical composition for the initial sphere before the collapse and we have much more observational constraints on the chemical composition of the cold core, i.e. objects with a density of a few $10^5$~cm$^{-3}$, rather than pre-stellar cores, i.e. objects with a density of a few $10^6$~cm$^{-3}$. We then assume that the transition between a dense core density and a pre-stellar core density is very rapid, affecting at most the depletion degrees of some species.
The species are assumed to be initially in an atomic form as in diffuse clouds except for hydrogen which is already molecular.
Elements with an ionization potential below the maximum energy of ambient UV photons ($13.6$~eV, the ionization energy of H atoms) are initially in a singly ionized state, i.e., C, S, Si, Fe, Na, Mg, Cl, and P.
We used the elemental abundances from \cite{hincelin_oxygen_2011}, with an oxygen elemental abundance equal to $1.5\times 10^{-4}$. The carbon-to-oxygen ratio of 1.13 in gas-grain modeling increases the overall agreement with observations of dark clouds \citep[see][]{hincelin_oxygen_2011}.
Table~\ref{tab:elem_ab} shows the elemental abundances used.
For each chemical element, we define the elemental abundance as the ratio of the number of nuclei of this element in the gas and in dust grain mantles to the total number of H nuclei.
This excludes the nuclei locked in the refractory part of the grains.
Thereafter, the abundance $A$ of a chemical species $\rm X$ refers to the number density $n$ of the considered species (located in the gas, in the grain mantles, or the sum of the two) relative to the total number density of hydrogen nuclei $n_{\rm H}$:
\begin{equation}
\label{eq:ab}
A({\rm X})=\frac{n({\rm X})}{n_{\rm H}}=\frac{n({\rm X})}{n({\rm H}) + 2n({\rm H_2})}.
\end{equation}

\begin{table}
\begin{center}
\caption{Elemental abundances relative to Hydrogen Nuclei}
\label{tab:elem_ab}
\begin{tabular}{lll}
\tableline
\tableline
Element & Abundance & Ref \\
\tableline
He & 9(-2) & (1) \\
C & 1.7(-4) & (2) \\
N & 6.2(-5) & (2) \\
O & 1.5(-4) & (3) \\
Na & 2(-9) & (4) \\
Mg & 7(-9) & (4) \\
Si & 8(-9) & (4) \\
P & 2(-10) & (4) \\
S & 8(-8) & (4) \\
Cl & 1(-9) & (4) \\
Fe & 3(-9) & (4) \\
\tableline
\end{tabular}
\end{center}
\tablecomments{$\rm a(b)$ means $a\times 10^b$}
\tablerefs{(1) see \cite{wakelam_polycyclic_2008}; (2) from \cite{jenkins_unified_2009}; (3) see \cite{hincelin_oxygen_2011}; (4) "low metal" values, see \cite{graedel_kinetic_1982}.}
\end{table}

\subsection{Interface between {\ttfamily{NAUTILUS}} and {\ttfamily{RAMSES}} codes}
\label{subsec:interface_nls_Ramses}

Using the computed chemical composition of the molecular cloud -- which has been run for $6\times 10^5$~yr -- as the initial condition, we then ran the chemistry during the collapse for $\sim 4\times 10^4$~yr depending on the model (see Table~\ref{tab:times_simulation}) with the three dimensional physical structure computed by {\ttfamily{RAMSES}}.
To do so, we included tracer particles in {\ttfamily{RAMSES}}.
They follow the fluid in movement and give us a set of one million trajectories per model.
Each trajectory comes along with the temperature and the density of the particle as a function of time.
{\ttfamily{NAUTILUS}} is then able to compute the chemical evolution using this information, assuming the gas and grain temperatures  are the same.
The method is described in detail in section~2.3 of HW13.

The central processing unit (CPU) time to compute the chemical evolution for one trajectory is about 30~s, which gives a total of roughly $10^4$~h for all the $10^6$ trajectories for each model using the JADE cluster from CINES.

\section{Results}
\label{sec:results}

\subsection{Physical characterization of the collapsing core}
We first present the physical characteristics of the collapsing core at the end of the simulation.
Then we present the past evolution of its temperature and its density.

\subsubsection{Characteristics of the components}
\label{subsubsec:components_charac}

\begin{figure}
\centering
\includegraphics[width=1.0\linewidth]{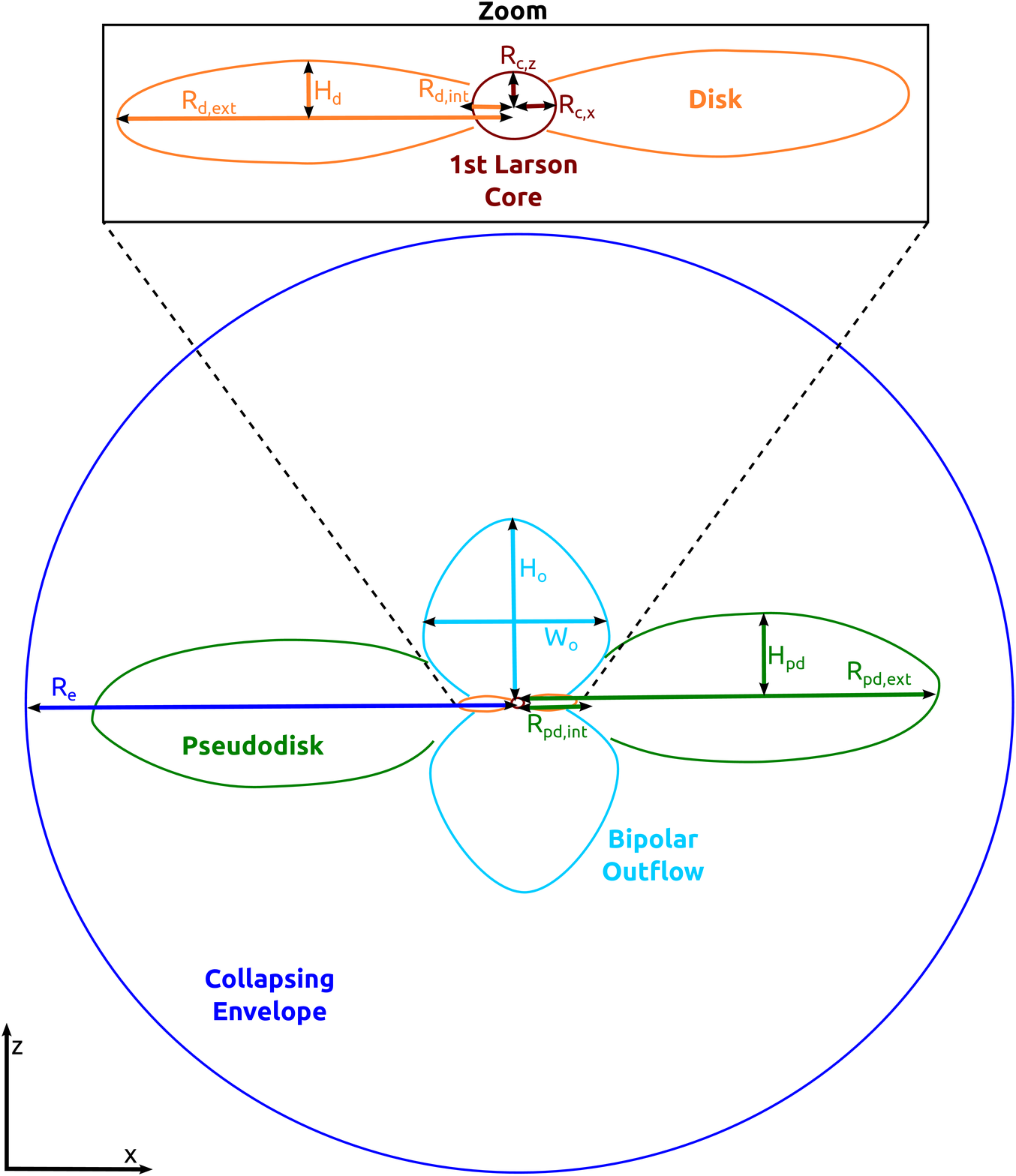}
\caption{Schematic view in the x-z plane of a collapsing core with its different components : central core (red), bipolar outflow (light blue), rotationally supported disk (orange), {magnetic pseudodisk} (green), and collapsing envelope (dark blue).
$\rm R_{c,x}$, and $\rm R_{c,z}$ are the radii of the central core respectively in the x-direction and the z-direction.
$\rm R_e$ is the radius of the collapsing envelope.
$\rm R_{d,int}$, $\rm R_{d,ext}$, and $\rm H_d$ are respectively the internal radius, the external radius, and the {height} of the rotationally supported disk.
$\rm R_{pd,int}$, $\rm R_{pd,ext}$, and $\rm H_{pd}$ are respectively the internal radius, the external radius, and the {height} of the {magnetic pseudodisk}.
$\rm H_o$ and $\rm W_o$ are respectively the height and the width of the bipolar outflow.
Values for these quantities are given in Table~\ref{tab:components_dimensions}.}
\label{fig:drawing_components}
\end{figure}

Within the structure of the collapsing core at the end of the simulation, we identify several components: 1) the central core which has become the first Larson core, 2) the bipolar outflow, 3) the rotationally supported disk, 4) {the magnetic pseudodisk}, and 5) the collapsing envelope, based on criteria given by \cite{joos_protostellar_2012}.
A schematic view of the different components of the collapsing core is presented in Figure~\ref{fig:drawing_components}.
Boundaries between pairs of components are not always noticeable.
Thus, we applied thresholds named $f_{\mathrm{thres}}$, $v_{\mathrm{thres}}$, $\rho_{\mathrm{thres}}$, and $T_{\mathrm{thres}}$ to the criteria that we describe below.

The central core is composed of particles for which the thermal support $e_\mathrm{i}$ is at least two times larger ($f_{\mathrm{thres}}=2$) than the rotational support $e_\mathrm{k}$.
The thermal support, or the density of internal energy, follows the equation
\begin{equation}
e_\mathrm{i}=\frac{1}{\gamma -1}\frac{\rho k_\mathrm{B} T}{\mu_{m} m_{\rm H}},
\end{equation}
where $\gamma$, $k_\mathrm{B}$, $\mu_m$, and $m_{\rm H}$ are respectively the adiabatic index, the Boltzmann constant, the mean molecular weight, and the mass of hydrogen nucleus.
The rotational support, or the density of rotational kinetic energy, satisfies
\begin{equation}
e_\mathrm{k}=0.5\rho v_\phi^2,
\end{equation}
where $v_\phi$ is the azimuthal velocity.
The azimuthal velocity of some particles is small enough to be selected in this component despite their low temperature (11 to 15~K) and their distance from the center of the system which runs from several hundreds to thousands of AU.
It is for this reason a minimum temperature and density are applied as supplementary criteria:
\begin{equation}
T > T_{\mathrm{thres}} = 100~{\rm K},
\end{equation}
\begin{equation}
\rho > \rho_{\mathrm{thres}}=10^9~{\rm cm^{-3}}.
\end{equation}
{$v_\phi$ is given in the frame of reference centered on the central core, so it is not suitable for the fragments surrounding the central core in the case of fragmentation.
Therefore, for the MU200$\Theta$0 model, we only apply the minimum temperature and density as criteria.}

{A particle that belongs to the bipolar outflow must have a velocity opposed to the collapse motion.
Therefore, we verify that the scalar product of the radial distance $\vec{r}$ and the velocity $\vec{v}$ is positive:
\begin{equation}
\vec{r}.\vec{v} = x v_x + y v_y + z v_z > 0,
\end{equation}
and we apply a velocity threshold to the radial velocity $v_r$:}
\begin{equation}
v_r > v_{\mathrm{thres}} = 0.2~{\rm km s^{-1}}.
\end{equation}

A particle that belongs to the rotationally supported disk satisfies the following criteria:
\begin{enumerate}
\item an azimuthal velocity $v_\phi$ more than two times larger than its radial velocity $v_r$ (the matter must not collapse too fast compared to its rotational motion);
\item an azimuthal velocity more than two times larger than its vertical velocity $v_z$ (this criteria removes the outflow cavity wall that is rotating fast);
\item a rotational support at least two times larger than the thermal support (to exclude the central core);
\item a density above a threshold value $\rho_{\mathrm{thres}}$ equal to $10^9$~cm$^{-3}$ to obtain more realistic estimates of the shape of the disk \citep[see][]{joos_protostellar_2012}.
\end{enumerate}

A particle that belongs to the {magnetic pseudodisk} satisfies previous criterion 3, but not criterion 1 or 2, and criterion 4 is relaxed to $10^7$~cm$^{-3}$.
{A few falling particles that are below or above the central core are also selected with these criteria, even though they are not a part of the pseudodisk.
Thus we apply an extra criterion to remove the majority of these particles in which the velocity $v_R$ (in cylindrical coordinates) has to be larger than the vertical velocity $v_z$ by a factor $f_{\mathrm{thres}}$:}
\begin{equation}
v_R = \sqrt{v_x^2+v_y^2} > f_{\mathrm{thres}} v_z.
\end{equation}

Finally, if a particle does not belong to one of the previous components, it is considered to  {{{be part of }}} the envelope, to which the following density and radius thresholds are also applied:
\begin{equation}
\rho < \rho_{\mathrm{thres}} = 10^7~{\rm cm^{-3}},
\end{equation}
\begin{equation}
r < r_0 \simeq 3300~{\rm AU}.
\end{equation}

\begin{table*}
\begin{center}
\caption{Number of selected tracer particles and {mass of components}.}
\label{tab:components_selection}
\begin{tabular}{lllllllllll}
\tableline
\tableline
Model & \multicolumn{2}{c}{Core} & \multicolumn{2}{c}{Outflow} & \multicolumn{2}{c}{Disk} & \multicolumn{2}{c}{Pseudodisk} & \multicolumn{2}{c}{Envelope} \\
          & Ratio & Mass & Ratio & Mass    & Ratio & Mass    & Ratio & Mass  & Ratio & Mass \\
\tableline
MU10$\Theta$0  & 6.2   & 0.05 & 2.8   & 3.5(-2) & 4.4   & 0.04    & 22    & 0.18  & 65    & 0.51 \\
MU10$\Theta$45 & 13    & 0.10 & 6.2   & 6.6(-2) & 8.5   & 0.06    & 17    & 0.08  & 55    & 0.44 \\
MU200$\Theta$0 & 25    & 0.19 &       &         & 5.7   & 0.05    & 15    & 0.12  & 54    & 0.43 \\
MU2$\Theta$0   & 4.7   & 0.05 & 0.3   & 3.3(-3) &       &         & 35    & 0.36  & 60    & 0.45 \\
\tableline
\end{tabular}
\tablecomments{For each component, the first column indicates the number of selected particles that belong to the corresponding component, {relative to the total number of selected particles of a given model, and the second column shows the mass of the component.
Calculation is done at the end of the simulations, at the times given in Table~\ref{tab:times_simulation}.
Ratio and mass are respectively given in \% and M$_\odot$.}
$\rm a(b)$ means $a\times 10^b$.}
\end{center}
\end{table*}

Table~\ref{tab:components_selection} shows the number of selected particles that belong to a given component, {relative to the total number of selected particles of a given model, in \%.
It also shows the corresponding mass of the components, in solar mass.}

MU2$\Theta$0 model does not have a disk due to the strong magnetic field.
This field produces a strong magnetic braking that decreases the azimuthal velocity of the matter, and so decreases the rotational support.
As a consequence, the pseudodisk is much larger for this model than for the MU10$\Theta$0 model.
Note the relatively low number of selected particles for the outflow component {(0.3~\%)} which comes from the powerful outflow due to the strong magnetic field, through magneto-centrifugal means \citep{blandford_hydromagnetic_1982,pudritz_centrifugally_1983,uchida_magnetodynamical_1985}.
{Few particles stay on this component, because the majority is quickly recycled in the envelope component.}

\begin{figure}
\centering
\includegraphics[width=1.0\linewidth]{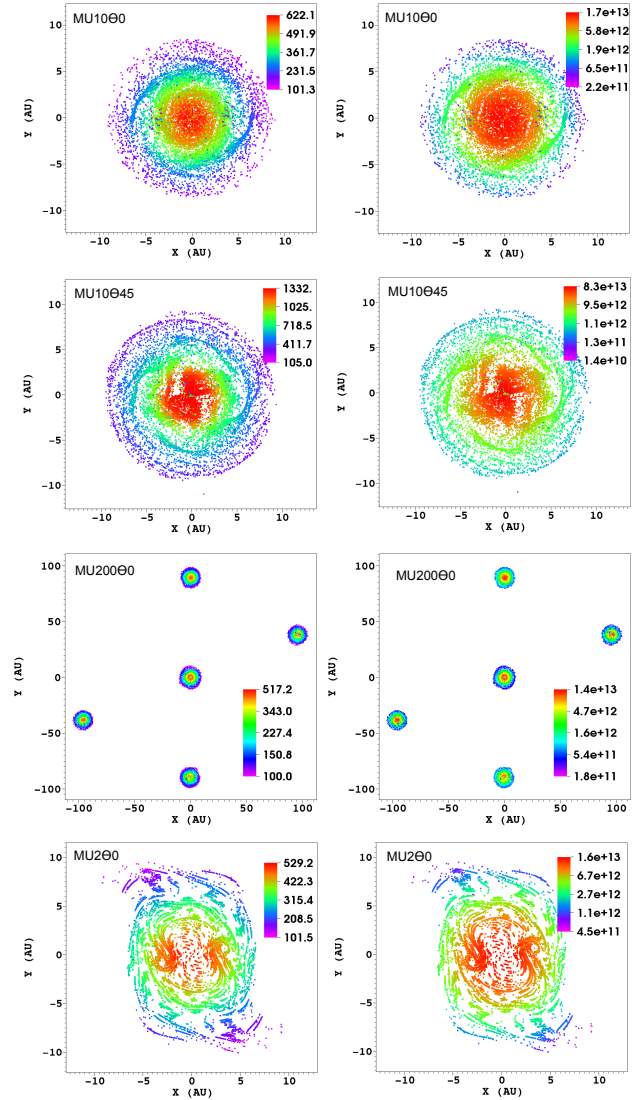}
\caption{Left and right panels respectively shows temperature (in K) and density (in cm$^{-3}$) of particles that belong to the core, at the end of simulations for MU10$\Theta$0, MU10$\Theta$45, MU200$\Theta$0, and MU2$\Theta$0 models.
Particles are projected onto the x-y plane.
Color coding for temperature is linear, and the one for density is logarithmic.
Color coding shows minimum and maximum values.}
\label{fig:temp_dens_core}
\end{figure}

MU200$\Theta$0 model has a weak magnetic field, so the outflow is absent.
For this model, more particles satisfy central core criteria.
Fragmentation has generated four hot spots around the central core as shown in Figures~\ref{fig:ramses_output} and \ref{fig:temp_dens_core}, and particles that are inside these hot spots are taken into account.

When the angle $\Theta$ between magnetic field lines and the rotation axis of the sphere is larger as in MU10$\Theta$45 model, the disk component is larger.
\cite{joos_protostellar_2012} show this general trend of an increase of the disk mass with the angle $\Theta$.
When this angle is lower, the magnetic braking is enhanced, which leads to a less massive disk \citep{hennebelle_disk_2009}.

\begin{table*}
\begin{center}
\caption{Size of components (AU).}
\label{tab:components_dimensions}
\begin{tabular}{lll|ll|lll|lll|l}
\tableline
\tableline
Model & \multicolumn{2}{c}{Core} & \multicolumn{2}{c}{Outflow} & \multicolumn{3}{c}{Disk} & \multicolumn{3}{c}{Pseudodisk} & Envelope \\
 & $\rm R_{c,x}$ & $\rm R_{c,z}$ & $\rm H_o$ & $\rm W_o$ & $\rm R_{d,int}$ & $\rm R_{d,ext}$ & $\rm H_d$ & $\rm R_{pd,int}$ & $\rm R_{pd,ext}$ & $\rm H_{pd}$ & $\rm R_{e}$ \\
\tableline
MU10$\Theta$0  & 8  & 6 & 250 & 360 & 9 & 90 -- 120\tablenotemark{a} & 30 & 60 -- 100 & 800 & 150 & 3300 \\
MU10$\Theta$45 & 9 & 8 & 450 & 600 & 9 & 140 -- 190 & 90 & 60 -- 80 & 660 & 220 & 3300 \\
MU200$\Theta$0 & 9 (9)\tablenotemark{b} & 6 (6) &  &  & 6 & 100 -- 200 & 30 & 40 -- 130 & 660 & 150 & 3300 \\
MU2$\Theta$0   & 9 & 4 & 420 & 440 &  &  &  & 10 & 1300 & 180 & 3300 \\
\tableline
\end{tabular}
\tablenotetext{1}{a -- b corresponds to minimum -- maximum.}
\tablenotetext{2}{Values between parenthesis correspond to the four "hot spots".}
\tablecomments{See Figure~\ref{fig:drawing_components} for more details.}
\end{center}
\end{table*}

\begin{table*}
\begin{center}
\caption{Temperature (K) and density (cm$^{-3}$) of the different components.}
\label{tab:components_temp_dens}
\begin{tabular}{llccc|ccc|ccc|ccc|ccc}
\tableline
\tableline
Model & Data & \multicolumn{3}{c}{Core} & \multicolumn{3}{c}{Outflow} & \multicolumn{3}{c}{Disk} & \multicolumn{3}{c}{Pseudodisk} & \multicolumn{3}{c}{Envelope} \\
 & & min & max & mean & min & max & mean & min & max & mean & min & max & mean & min & max & mean \\
\tableline
MU10$\Theta$0 & T & 101 & 622 & 431 & 11 & 29 & 12 & 11 & 160 & 17 & 11 & 115 & 11 & 11 & 11 & 11 \\
 & n & 2(11) & 2(13) & 7(12) & 9(6) & 1(10) & 2(8) & 1(9) & 7(11) & 1(10) & 1(7) & 3(11) & 2(8) & 6(4) & 1(7) & 9(5) \\
\tableline
MU10$\Theta$45 & T & 105 & 1332 & 1179 & 11 & 56 & 12 & 12 & 333 & 26 & 11 & 35 & 12 & 11 & 12 & 11 \\
 & n & 1(10) & 8(13) & 5(13) & 4(6) & 4(9) & 7(7) & 1(9) & 5(11) & 4(9) & 1(7) & 6(9) & 9(7) & 7(4) & 1(7) & 7(5) \\
\tableline
MU200$\Theta$0 & T & 100 & 517 & 285 & & & & 11 & 100 & 23 & 11 & 100 & 14 & 11 & 11 & 11 \\
 & n & 2(11) & 1(13) & 4(12) & & & & 1(9) & 6(11) & 2(10) & 1(7) & 6(11) & 3(8) & 5(4) & 1(7) & 8(5) \\
\tableline
MU2$\Theta$0 & T & 102 & 529 & 377 & 11 & 261 & 43 & & & & 11 & 241 & 12 & 11 & 12 & 11 \\
 & n & 5(11) & 2(13) & 8(12) & 8(5) & 3(12) & 1(10) & & & & 1(7) & 2(12) & 1(8) & 6(4) & 1(7) & 8(5) \\
\tableline
\end{tabular}
\tablecomments{$\rm a(b)$ means $a\times 10^b$.
Column named "mean" gives 10\textasciicircum$\left({\frac{1}{N}\sum\limits_{p=1}^N\log T(p)}\right)$ and 10\textasciicircum$\left({\frac{1}{N}\sum\limits_{p=1}^N\log n(p)}\right)$, where $N$ is the number of particles that belong to the considered component.
Values are spread over multiple orders of magnitude, thus the logarithm is used to calculate the averaged value.}
\end{center}
\end{table*}

\begin{figure}
\centering
\includegraphics[width=1.0\linewidth]{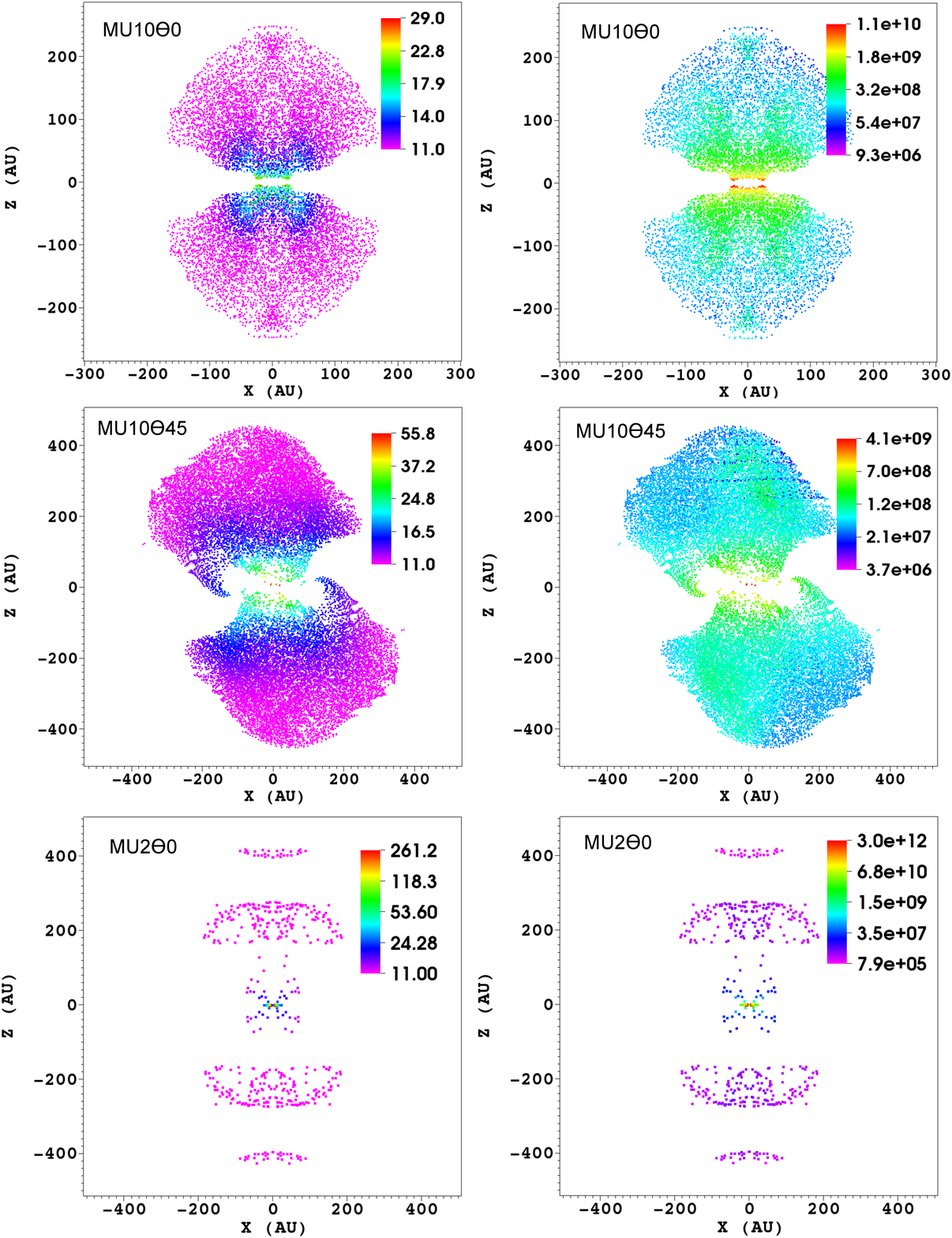}
\caption{{Same as figure~\ref{fig:temp_dens_core} but for the outflow.
Particles are projected onto the x-z plane.
Model MU200$\Theta$0 does not have an outflow.}}
\label{fig:temp_dens_outflow}
\end{figure}

\begin{figure*}
\centering
\includegraphics[width=1.0\linewidth]{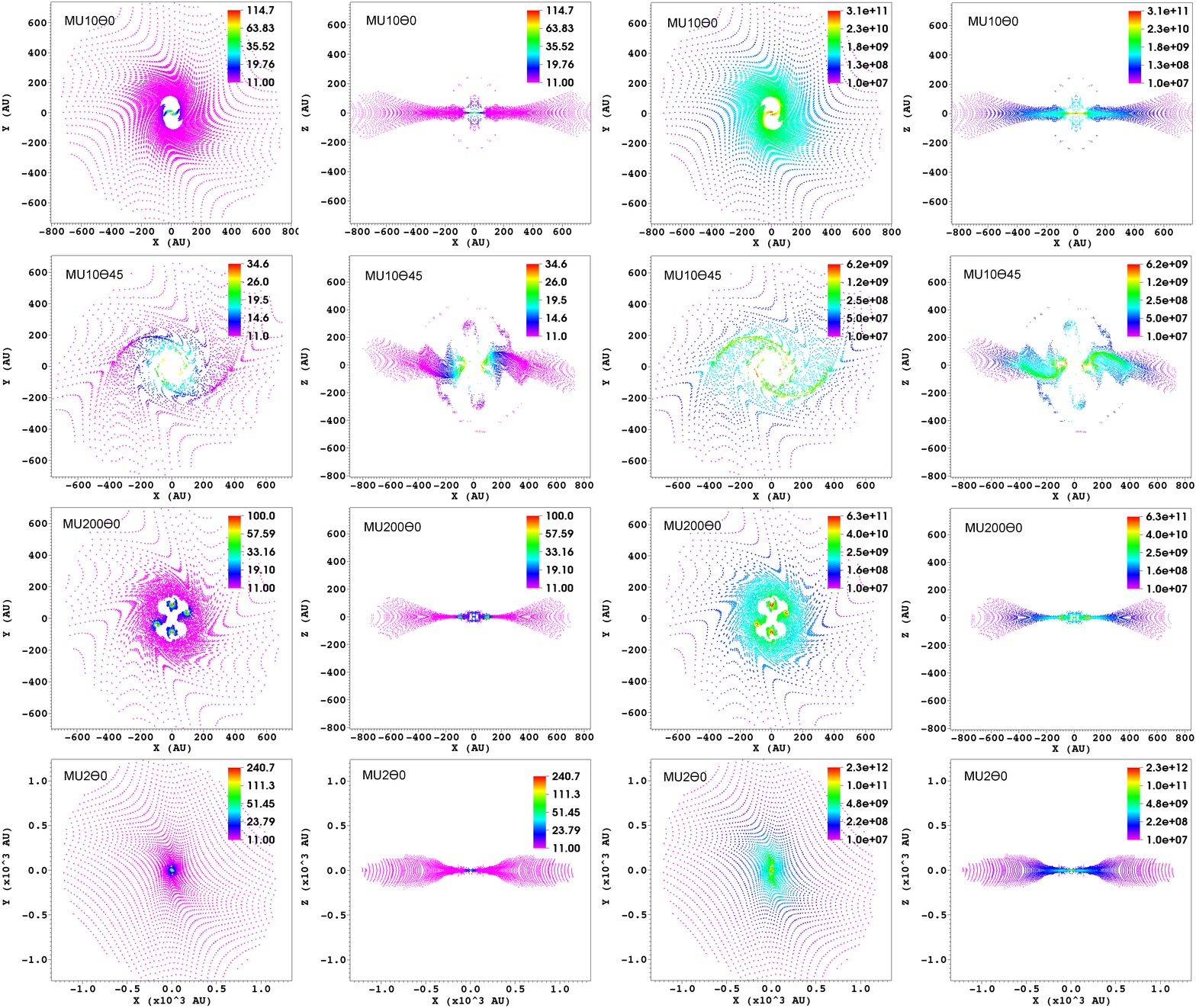}
\caption{{Same as figure~\ref{fig:temp_dens_core} but for the pseudodisk.
Particles are projected onto the x-y plane with a selective threshold $|z|<5$~AU in the first and third columns, and onto the x-z plane with a selective threshold $|y|<50$~AU in the second and fourth columns.}}
\label{fig:temp_dens_pseudodisk}
\end{figure*}

\begin{figure*}
\centering
\includegraphics[width=1.0\linewidth]{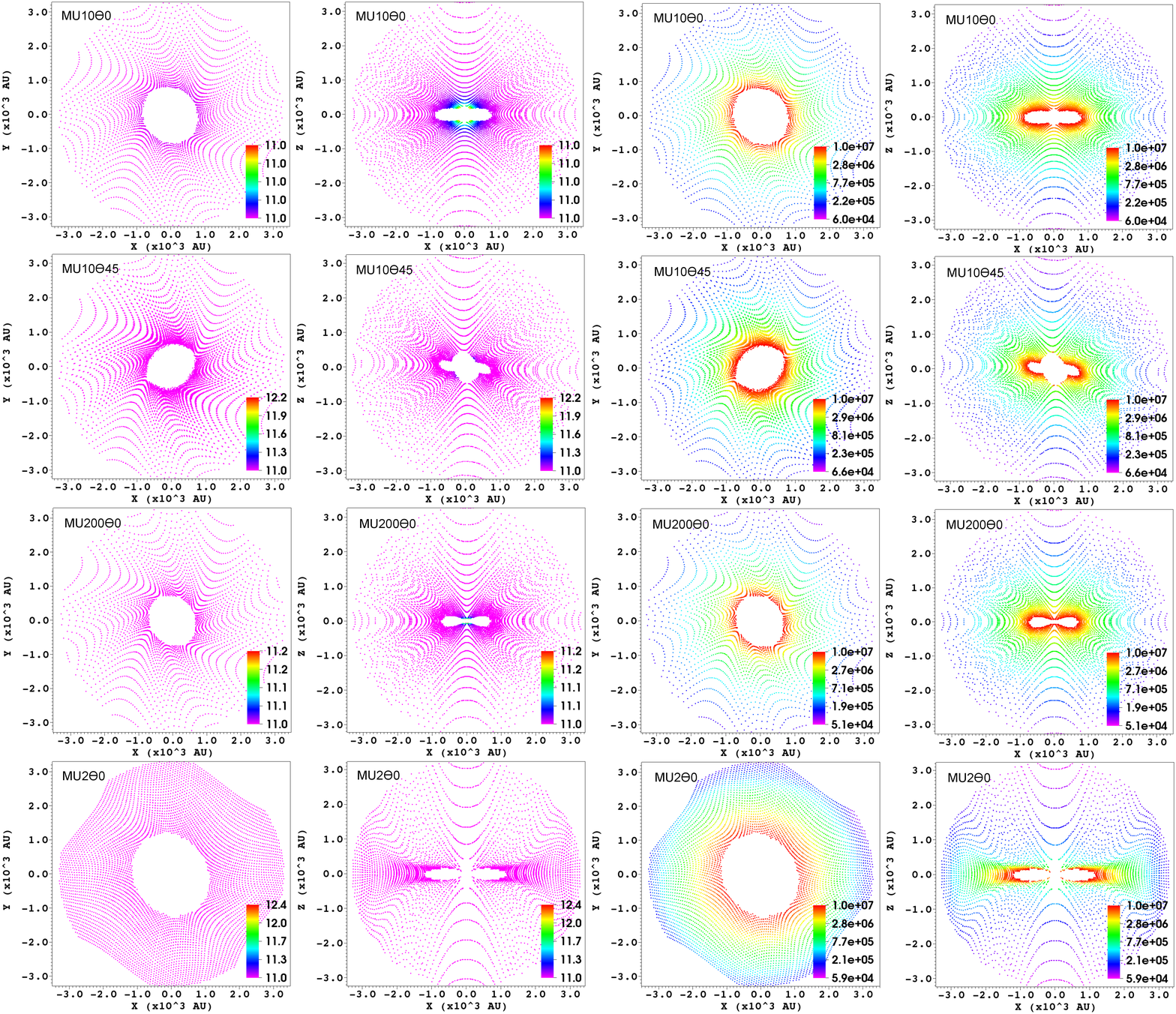}
\caption{{Same as figure~\ref{fig:temp_dens_core} but for the envelope.
Particles are projected onto the x-y plane in the first and third columns, and onto the x-z plane in the second and fourth columns, with a respective selective threshold $|z|<50$~AU and $|y|<50$~AU.}}
\label{fig:temp_dens_envelope}
\end{figure*}

Table~\ref{tab:components_dimensions} gives the size of the components and refers to Figure~\ref{fig:drawing_components}, and Table~\ref{tab:components_temp_dens} gives their temperature and density.
Figures~\ref{fig:temp_dens_core} to \ref{fig:temp_dens_envelope} show a two dimensional view of the components.
These figures display the temperature (on the left panel) and density (on the right panel) of particles that belong to the components.
The temperature and the density of the disk component are given in  Figures~2 and 3 by HW13.
From a general point of view, the matter is relatively cold in the majority of the collapsing core, and the temperature rises abruptly in the central core.
Density values are spread over multiple orders of magnitude, however, on a larger spatial scale.
For example, in the MU10$\Theta$0 model, the temperature is lower than 20~K at 40~AU from the central core and beyond, and reaches around 600~K in the central core.
In the center, the density is around $10^{13}$~cm$^{-3}$, while at 10~AU, 100~AU, and 1000~AU, it reaches respectively around $10^{11}$~cm$^{-3}$, between $10^{11}$~cm$^{-3}$ and $10^9$~cm$^{-3}$ depending on the x-y-z position, and $10^7$~cm$^{-3}$.  
Numbers presented in the following (minimum, maximum, and mean density and temperature, and size) certainly depend on the values adopted for our criteria.
However, slightly changing these values will not significantly affect these numbers.

Central cores are displayed in  Figure~\ref{fig:temp_dens_core}.
They have similar sizes for the different models.
Their equatorial radii, around 9~AU, are larger than their vertical radii, around 6~AU.
This slightly flattened oblate shape comes from {{{the remainder of the initial}}} rotation of the system.
Temperature and density ranges, from 100~K and $10^{10}$~cm$^{-3}$ to 600~K and $10^{13}$~cm$^{-3}$, are similar for all models except for MU10$\Theta$45 model which presents a hotter and denser center.
Note the four hot spots in the MU200$\Theta$0 model due to fragmentation processes, which have similar density and temperature as the central core.
Our calculations are stopped before the onset of the second collapse and before dust grain destruction, which explains the relative low central temperature we obtain.

Outflows are displayed in  Figure~\ref{fig:temp_dens_outflow}.
Due to the strong magnetic field strength, the outflow from the MU2$\Theta$0 model has the highest velocity which is $\sim 4$~km s$^{-1}$ (see Figure~\ref{fig:ramses_output}).
Note that this value is quite low compared to the velocity observed in low-mass protostars, which is around 10--100~km s$^{-1}$ \citep[see for example][]{arce_molecular_2007}.
The outflow observed in our models corresponds to a low-velocity component driven at the FHSC scale.
The high-velocity component is driven at much smaller scales during the second collapse \citep[e.g.,][]{machida_high-_2008}, but we stop our calculations before the onset of the second collapse.
The outflow of the MU10$\Theta$0 and MU10$\Theta$45 models, which have a moderate magnetic field strength, have velocities around 1--2~km~s$^{-1}$.
The outflows are mostly cold, with an average temperature around 12~K, but {present temperatures up to about 60~K} close to the central core.
{The outflow of the MU2$\Theta$0 model is however hotter due to some selected particles very close to the central core.}
The density in these outflows shows large variations, especially for the MU2$\Theta$0 model (from $8\times 10^{5}$ to $3\times 10^{12}$~cm$^{-3}$).
The MU200$\Theta$0 model does not possess any outflow because the magnetic field strength is very low.
The high density and cold temperature observed in the outflow can be explained by the following.
The collapsing core is embedded and the outflow is young, so the cavity formed is not as "empty" as in older outflows such as observed in Class~0 protostars \citep{gomez-ruiz_density_2015}.
As a consequence, optical depth is important and the matter stays mostly cold.
Also, the central core does not irradiate as strongly as a Class~0 protostar, so it does not warm the surrounding as much.
A few selected particles are very close to the central core, at the interface between the outflow and the central core, which explains the high maximum value of the density we obtain.
These particles are not numerous so they do not change  the mean value of the density and the temperature significantly.

Disks are displayed in  Figures~2 and 3 from HW13.
The disk component is not present in every model as mentioned previously, and can have a very different shape as a function of the model.
The MU2$\Theta$0 model has a strong magnetic field which prevents the formation of a rotationally supported disk while the MU200$\Theta$0 model presents a fragmented disk.
However, a {disk-like} shape component with a radius equal to $\sim$40~AU is observed around the central core of the MU200$\Theta$0 model.
The MU10$\Theta$45 model shows a warped disk, due to the angle $\Theta$ between magnetic field lines and the rotation axis of the sphere.
Spiral arms can be seen in the disks of the MU10$\Theta$0 and MU10$\Theta$45 models.
In general, disks possess relatively large variations of temperature and density conditions {from 11~K and $10^{9}$~cm$^{-3}$ in the outer radius to about 100--330~K and $10^{11}$--$10^{12}$~cm$^{-3}$ in the inner radius.}
More details on the disk component are given in HW13.

Pseudodisks are displayed in Figure~\ref{fig:temp_dens_pseudodisk}.
The pseudodisk component is much larger than the disk component.
While the disk one has an external radius around 100--200~AU, the radius of the pseudodisk component {is about 700 to 1300~AU}.
There is an overlap between the external radii of disks and the internal radii of pseudodisks.
In this particular region, the matter begins to be rotationally supported against the collapse toward the central core.
The pseudodisk of the MU2$\Theta$0 model is the largest one -- its external radius is about 1300~AU, compared to 700 to 800~AU for the other models -- and shows the largest extrema for density and temperature values: from 11~K and $10^{7}$~cm$^{-3}$ to 241~K and $2\times 10^{12}$~cm$^{-3}$.
{However, the averaged temperature and density are similar for all models, respectively 11 to 14~K and $9\times 10^7$ to $3\times 10^{8}$~cm$^{-3}$.}
Despite the fact that MU200$\Theta$0 does have a weak magnetic field strength, a group of particles are selected according to the above criteria of a pseudodisk, {which is however not magnetic}.

Collapsing envelopes are displayed in  Figure~\ref{fig:temp_dens_envelope}.
Collapsing envelopes of all models are cold and quasi-isothermal ($\simeq 11-12$~K), while their density varies between $5\times 10^{4}$~cm$^{-3}$ in the outer regions to $10^{7}$~cm$^{-3}$ in the inner regions, with an average around $8\times 10^{5}$~cm$^{-3}$.
These low temperatures indicate that the collapsing envelope is not at the hot corino stage, with temperatures of 100~K and intensive rotational emission from complex organic molecules.

\subsubsection{Physical history of the matter}
\label{subsubsec:phys_history_matter}

\begin{figure*}
\centering
\includegraphics[width=0.75\linewidth]{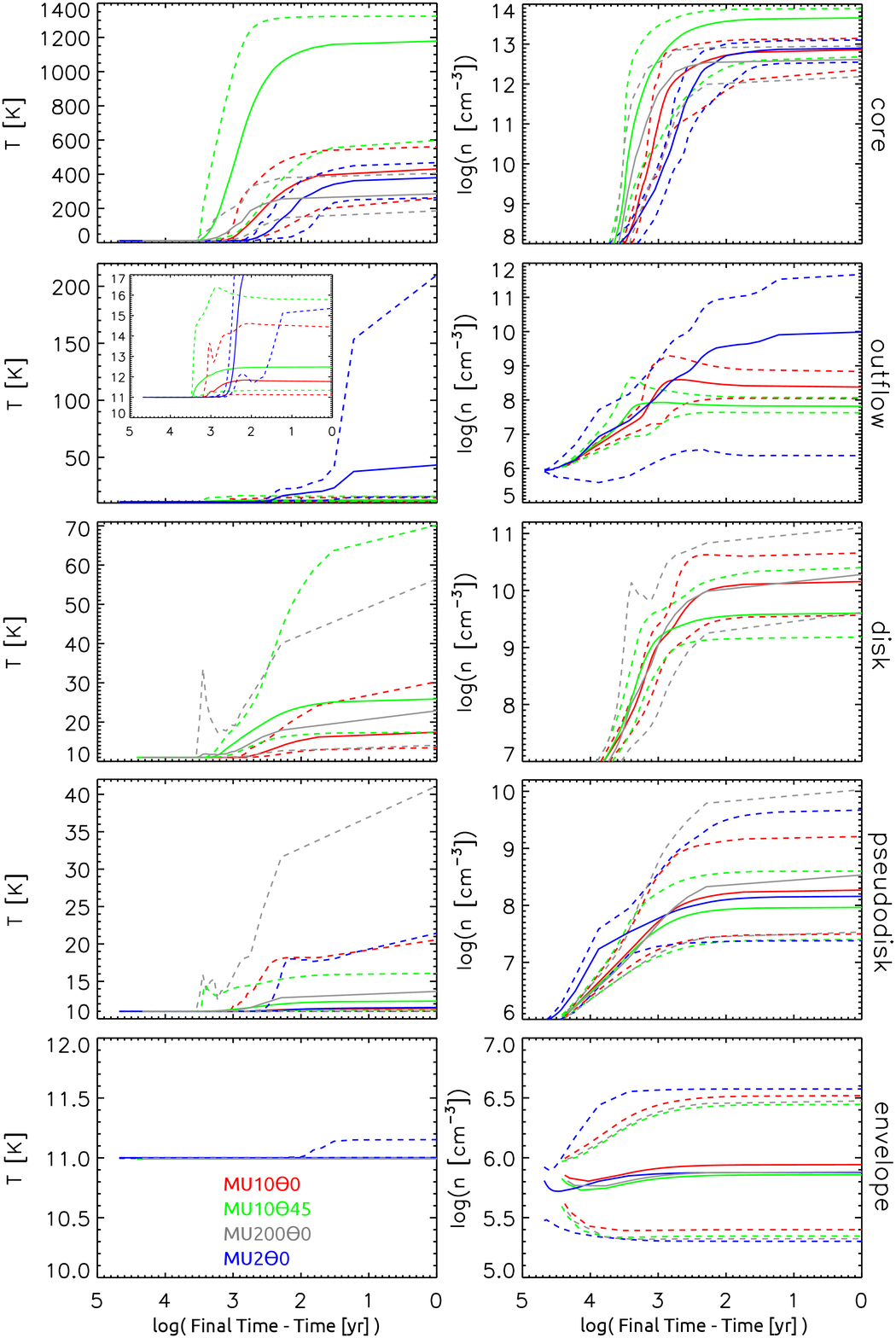}
\caption{Evolution of mean temperature and mean density of components during the collapse (solid lines), with associated standard deviation (dotted lines), for MU10$\Theta$0 (red), MU10$\Theta$45 (green), MU200$\Theta$0 (grey), and MU2$\Theta$0 (blue) models.
The abscissa displays the logarithm of time remaining until the final time.
See text for details on mean and standard deviation calculations.}
\label{fig:history_models}
\end{figure*}

Figure~\ref{fig:history_models} presents the evolution of the mean temperature and the mean density of the matter that constitutes each component at the final time for the four models.
As mentioned in the note of  Table~\ref{tab:components_temp_dens}, the mean quantity $\left\langle \log X(t)\right\rangle$ follows the equation 
\begin{equation}
\label{eq:mean}
\left\langle \log X(t)\right\rangle = \frac{1}{N}\sum\limits_{p=1}^N\log X_p(t),
\end{equation}
where $t$ is the time, $N$ is the number of particles that belong to the considered component at the final time, and $p$ is a particle with $\log X_p$ the decimal logarithm of its temperature or its density.
The computed standard deviation $\sigma\left(t \right)$, as a function of time $t$, follows the equation 
\begin{equation}
\sigma\left(t \right) = \sqrt{\frac{1}{N}\sum\limits_{p=1}^N\left(\log X_p\left(t\right) - \left\langle \log X\left(t\right)\right\rangle\right)^2}.
\end{equation}
Two different values of the standard deviation are necessary to characterize the distribution, which is not symmetric about the mean:
one if $\log X_p\left(t\right) - \left\langle \log X\left(t\right)\right\rangle \geq 0$, and another if $\log X_p\left(t\right) - \left\langle \log X\left(t\right)\right\rangle < 0$.

A general trend distinguishes the temperature and the density evolutions along trajectories.
While temperature stays roughly constant (at 11~K) until around 1000~yrs before the end of the simulations, the density rises gradually {-- on a logarithmic scale --} from $\sim10^6$~cm$^{-3}$ at the start of the collapse, up to $\sim10^{13}$~cm$^{-3}$ at the end of the simulation depending on the considered component.
We will see that this trend has a strong consequence on the chemical evolution; molecules are first adsorbed on the grain surface {-- the higher the density, the faster this process --} and eventually react on the surface, before to eventually desorbing if the temperature is high enough.

From a general point of view, the physical history of the matter depends on the considered component.
While the physical characteristics of the matter that constitutes envelopes globally do not evolve much as a function of time,
 those that constitute central cores show huge variations.
They are the two extreme cases.
{If we focus on the range of values as a function of time,} the matter of the outflows, disks, and pseudodisks show some noticeable differences.
{The range of temperature of the disks at a given time is often larger than that of outflows and pseudodisks, while the range of density of the pseudodisks is often larger than that of other components.}
On the one hand, the disk is a very inhomogeneous component, with a high temperature in the inner region due to the proximity to the central core, and a cold outer region.
This  dichotomy leads to a higher standard deviation for the thermal history of this component compared to the other components.
On the other hand, the pseudodisk is an extended component and mainly cold.
This spatial extension leads to an important deviation in the density values.
There is a characteristic feature in the outflow history, particularly visible for MU10$\Theta$0 model in which a small peak occurs in temperature and density around 1000~yrs before the end of the simulations.
At this time, some of the matter comes relatively close to the hot and dense center of the collapsing core, and then is trapped in the outflow and as a consequence is taken away from the center.
We also observe a characteristic feature in the disk and the pseudodisk history of the MU200$\Theta$ model.
Temperature and density increase for a group of particles between around $10^4$ and $10^3$~yr before the end of the simulation.
The number of particles from this group is small enough to not impact the mean evolution of the temperature and the density, but does change  the standard deviation significantly.
For this model, the core is fragmented.
During that range of time -- $10^4$ and $10^3$~yr before the end of the simulation -- the particles come close to the fragments, which enhances the temperature and the density.
The particles do not bind to the fragments, however,  and finally will constitute the disks and the pseudodisk.

\begin{figure}
\centering
\includegraphics[width=1.0\linewidth]{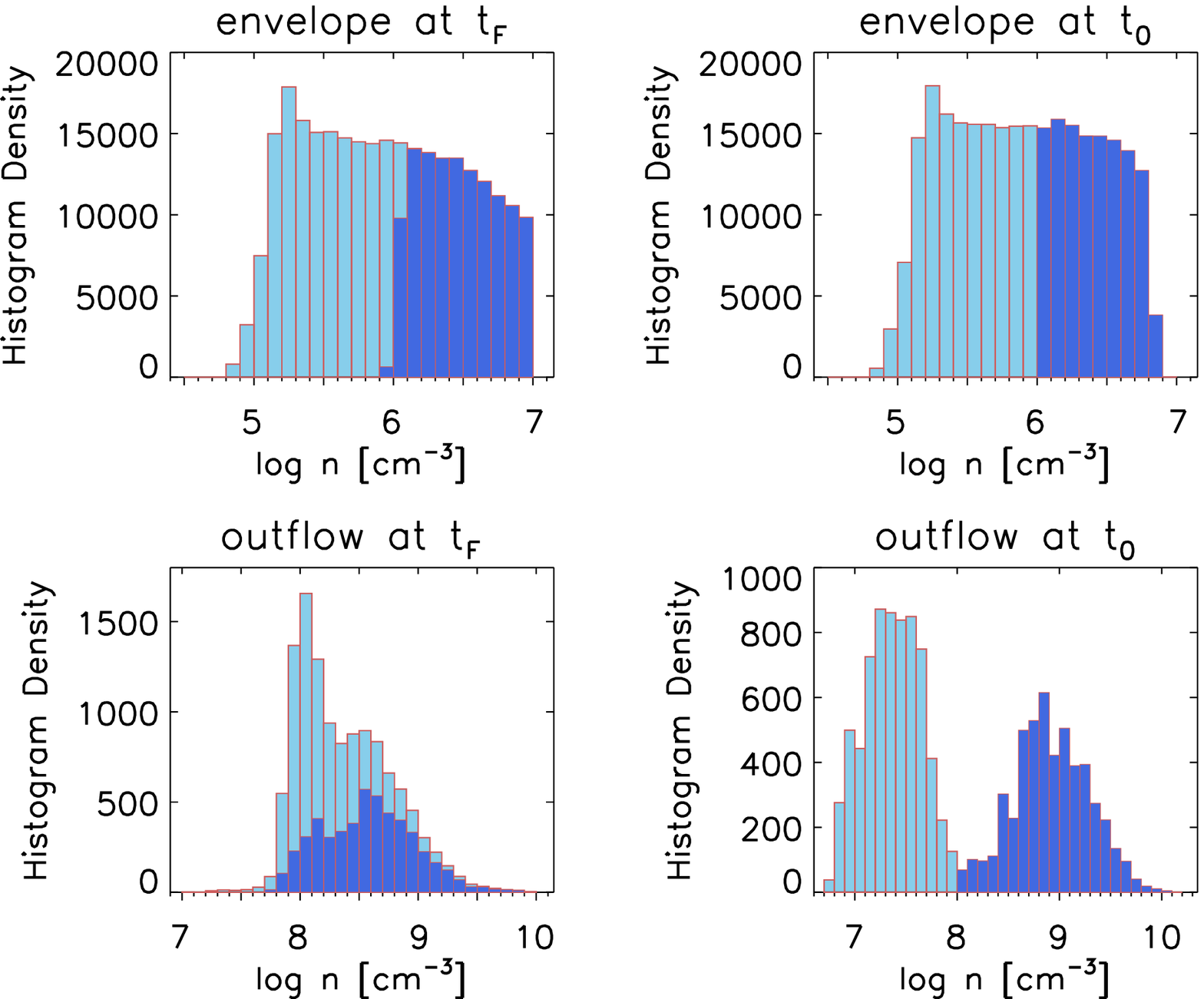}
\caption{Distribution of density of particles that belong at the end of the simulations to the envelope (top) and the outflow (bottom) for the MU10$\Theta$0 model.
Left and right panels respectively correspond to the final time (t$\rm _F$) and to $t_0$ (see text).
Light blue and dark blue distributions do not overlap but are accumulated.
Distributions for temperature and other components and models are given in Appendix~\ref{appendix:histogram}.}
\label{fig:histogram_example}
\end{figure}

Although these mean evolutions give useful global information on the history of the components, they do not show possible disparities among the group of particles that belong to a given component.
{Figure~\ref{fig:histogram_example} presents distributions of the density of particles that belong to the envelope and the outflow, both at the final time and at $t_0$ for the MU10$\Theta$0 model.
Regarding the envelope, the distribution of density values is relatively well reproduced by the mean density, for both times while for the outflow, this is also true at the final time, but at $t_0$ two different groups of particles appear (highlighted in light and dark blue).
This pattern indicates that these groups of particles, with a different physical history, have been gathered to form one final component.
A similar separation procedure is computed for the envelope, but we see that the "mixing" is very limited.
Figures~\ref{fig:appendix_histogram_mu10theta0} to \ref{fig:appendix_histogram_mu2theta0} in Appendix~\ref{appendix:histogram} present the results for all components and models.
The conclusion is similar for most of the envelopes, and the other components often present a more complex history.
On the one hand, this result shows a limit to the usefulness of the mean evolution for some components and models.
On the other hand, it reveals the usefulness of multi-dimensional simulations, in order to take into account the complex evolution of the spatial distribution of the matter.}

The mean density of the envelope does not evolve much as a function of time, but we emphasize  the fact that the dynamics are essential to obtain a realistic distribution of the matter that constitutes the envelope at the early stage of the simulation.
We start from a homogeneous distribution with a density of about $10^6$~cm$^{-3}$, and in less than $10^4$~yr, the density spreads within the range $5\times 10^4$ to $10^7$~cm$^{-3}$ in a Bonnor-Ebert like sphere.
Note that the initial homogeneous distribution of the density of the core is modified to a more realistic distribution quickly enough so that the chemistry is not significantly dependent on the initial density profile.  See Section~\ref{subsec:timescales} for more detail on the dynamical and the chemical timescales.

\subsection{Chemical characterization of the collapsing core}
\label{subsec:chem_charac}

\begin{figure}
\centering
\includegraphics[width=0.75\linewidth]{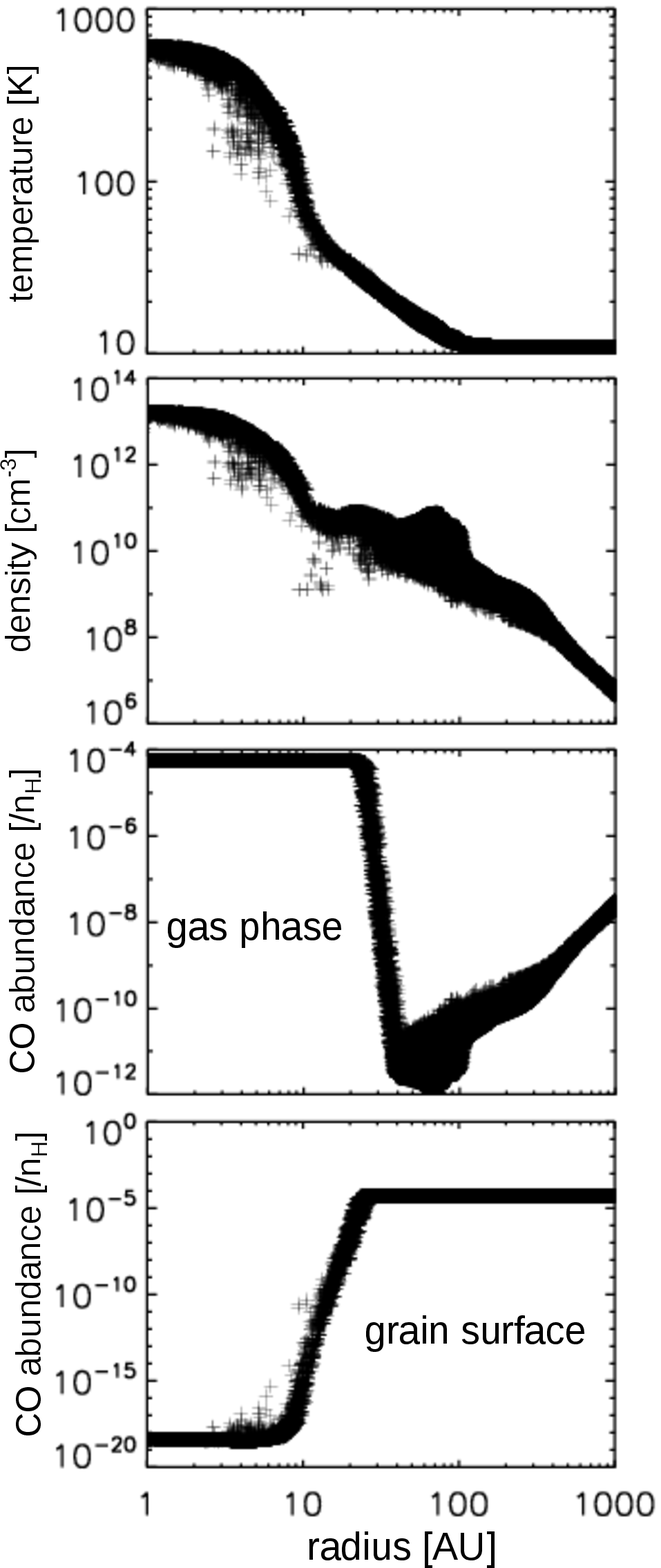}
\caption{Temperature, density, and CO abundances of particles that belong to the equatorial midplane (where $|z|<10$~AU) as a function of the radius, at the end of the MU10$\Theta$0 model simulation.
The density dispersion (from $10^9$ to $10^{11}$~cm$^{-3}$) between around 20 and 100~AU is due to the spiral arms, where the density is locally higher than the surrounding area, and has a direct consequence on the gas phase abundances through adsorption on grain surfaces.}
\label{fig:temp_dens_CO_midplane_mu10theta0}
\end{figure}

Due to the high density of the collapsing core, from $~10^5$ to $~10^{14}$~cm$^{-3}$ depending on the zone, and the relative short timescale of the collapse, less than $10^5$~yrs, the main processes occurring are the gas-grain interactions, namely the adsorption of molecules on the grains surface, and the desorption of these molecules from the grains.
Chemical reactions can however become important close to the central core, where the temperature is higher.
Figure~\ref{fig:temp_dens_CO_midplane_mu10theta0} shows the CO abundance as a function of the distance from the central core, in the midplane of the system ($|z|<10$~AU), at the end of the MU10$\Theta$0 simulation.
This figure illustrates a general trend concerning the chemical abundances as a function of the radius, due to the gas-grain interaction.
Gas-phase abundances decrease gradually from the outer part to the inner part of the collapsing core, as the density increases, i.e. as adsorption is enhanced.
This is true until the temperature becomes high enough to allow desorption of molecules from the grain surface.
At this point, gas-phase abundances are enhanced and the ices on grain surface are sublimated.
To determine if either adsorption or desorption dominates, we can compute and compare their timescales.
Appendix~\ref{appendix:timescales} gives a detailed calculation of theses quantities.
Table~\ref{tab:ads_des_timescales} shows the desorption and adsorption timescales of the CO molecule, for different temperature and density values, obtained by this calculation.
When the desorption timescale becomes larger than the adsorption timescale, ices are efficiently sublimated.
This transition occurs around 25~AU from the central core for CO, as seen in Figure~\ref{fig:temp_dens_CO_midplane_mu10theta0}, and is confirmed by Table~\ref{tab:ads_des_timescales} comparing adsorption and desorption timescales.

Non-thermal desorption also occurs.
Relativistic Fe nuclei, which are part of the stream of cosmic rays, can impulsively heat grains, and so induce desorption \citep{leger_desorption_1985}.
However, its timescale~\footnote{The non-thermal desorption timescale is equal to $\frac{1}{k_\mathrm{crd}}$, where $k_\mathrm{crd}$ is the desorption rate induced by cosmic rays \citep[see][]{hasegawa_new_1993}.} is larger than the collapsing timescale, thus it is less critical than thermal desorption.
For CO, the non-thermal desorption timescale is about $1.3\times 10^6$~yrs.
Other non-thermal desorption mechanisms may also occur such as reactive desorption and photodesorption.
Reactive desorption is set to 1~\% in our modeling, and occurs for every exothermic  one-product surface reaction.
The efficiency of this mechanism on icy grains is still uncertain \citep{minissale_influence_2014} but it seems to be higher for small molecules with low binding energies \citep{minissale_dust_2015}.
We use the "classic" 1~\% value, which is sufficient to reproduce the observation of methanol in cold cores \citep{garrod_non-thermal_2007,vasyunin_reactive_2013,minissale_dust_2015}.
Photodesorption follows \cite{oberg_photodesorption_2007} and \cite{hassel_modeling_2008}.
This mechanism is limited due to the high visual extinction.

\begin{figure}
\centering
\includegraphics[width=1.0\linewidth]{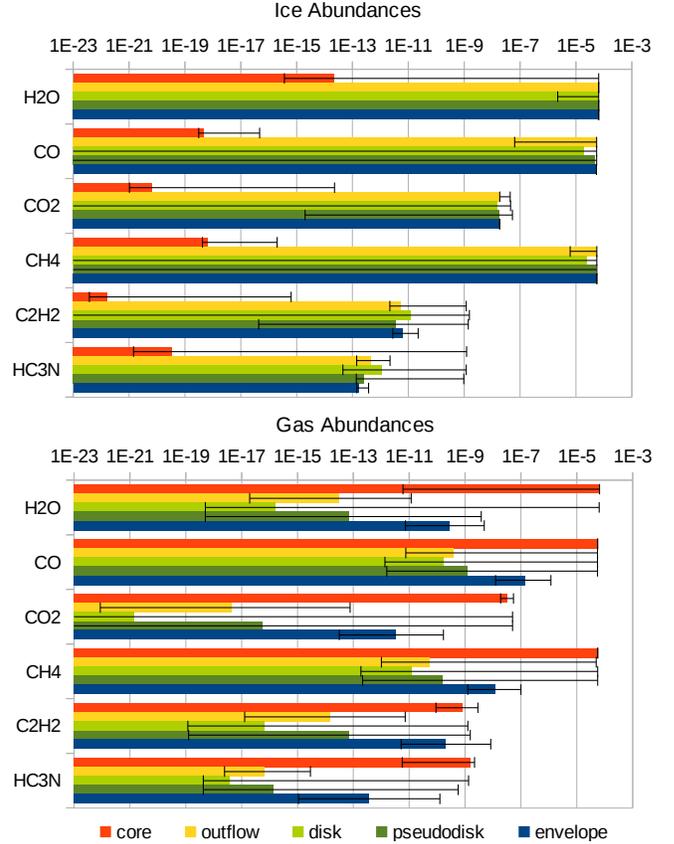}
\caption{Mean abundances of a selection of species, relative to the total density of hydrogen nuclei, on the grain surface (top), and in the gas phase (bottom), for the MU10$\Theta$0 model at the end of the simulation.
Abundances are averaged over the number of tracer particles that belong to a given component (core, outflow, disk, pseudodisk, or envelope) -- see Footnote~\ref{fnote:mean_abundance}.
Black lines show ranges of abundance values, for a given species in a specific component.}
\label{fig:mean_ab_ice_gas_mu10}
\end{figure}

These gas-grain interactions have important consequences on the relative contents between gas phase and grain surface for the different components of a given model.
Figure~\ref{fig:mean_ab_ice_gas_mu10} shows the mean abundance\footnote{\samepage
\label{fnote:mean_abundance}
We apply Equ.~\ref{eq:mean} with $X \equiv A({\rm CO})$.
$A({\rm CO})$ follows Equation~\ref{eq:ab}.
}
of a selection of species, in the gas phase and on the grain surface, for the core, the outflow, the disk, the pseudodisk, and the envelope of the MU10$\Theta$0 model.
While the mean ice abundances of a species is often similar for all components, except the central core where ices are sublimated, the gas phase abundances strongly depend on the component.
{Ice abundances do vary, but are less sensitive than gas phase abundances, because they are often the reservoirs.}
Abundances values depend on species, but a similar pattern can be seen for all species.
Mean values are maximal in the central core, then we observe a gradual depletion as a function of the component: the envelope, where the depletion is low, then the pseudodisk, the outflow, and finally the disk, where the depletion is the highest.
This behavior is a direct consequence of the amplitude of the gas-grain interaction, which is enhanced when the density is higher, until the temperature is high enough to allow thermal desorption.
Figure~\ref{fig:mean_ab_ice_gas_mu10} also shows the different size of abundances ranges, as a function of species and component.
Generally, ranges are small in the envelope, because this component is homogeneous in term of temperature and density conditions, while ranges are big in the disk and the pseudodisk because these components present inhomogeneity, due for example to the spiral arms where the density is very high and the temperature is low, or to regions close to the core where the temperature is high.
These ranges depend also on the species because they do not have the same desorption energy.

\begin{table*}
\begin{center}
\caption{Characteristic timescales for adsorption and desorption of CO, as a function of the density and the temperature.}
\label{tab:ads_des_timescales}
\begin{tabular}{cccc|cc}
\tableline
\tableline
Gas & Gas & Adsorption & Corresponding & Grain & Desorption \\
temperature & density & timescale & radius in Figure~\ref{fig:temp_dens_CO_midplane_mu10theta0} & temperature & timescale \\
$\rm [K]$ & $\rm [cm^{-3}]$ & & $\rm [AU]$ & $\rm [K]$ \\
\tableline
10 & 1(6)  & 3.6(3)~yr & 2000 & 10 & 2.7(30)~yr \\
10 & 1(10) & 4.4~months & 100 & 10 & 2.7(30)~yr \\
15 & 1(10) & 3.6~months & 60 & 15 & 6.2(13)~yr \\
20 & 1(10) & 3.1~months & 40 & 20 & 2.9(5)~yr \\
25 & 3(10) & 28~days & 26 & 25 & 3~yr \\
30 & 3(10) & 26~days & 20 & 30 & 12~h \\
35 & 5(10) & 24~days & 16 & 35 & 3~min \\
40 & 5(10) & 22~days & 14 & 40 & 3~s \\
50 & 1(11) & 20~days & 11 & 50 & 10~ms \\
60 & 1(11) & 18~days & 10 & 60 & 0.2~ms \\
\tableline
\end{tabular}
\tablecomments{$\rm a(b)$ means $a\times 10^b$. A detailed calculation of timescales is given in appendix~\ref{appendix:timescales}.
Each gas temperature and density pair roughly corresponds to a specific radius in Figure~\ref{fig:temp_dens_CO_midplane_mu10theta0}.
Sublimation of CO ices is then observed around 25~AU from the central core.}
\end{center}
\end{table*}

\subsubsection{Distribution of the molecules}
\label{subsubsec:distri_molecules}

We now describe the distribution of the main molecules among the core, the outflow, the disk, the pseudodisk, and the envelope of the models.
In this section, total abundances refer to the total content of species present in the gas phase and on the grain surface.

Figure~4 of HW13 shows the starting total abundances in the initial cloud, that can be compared to the following total abundances.
Chemistry occurring during the cloud phase, before the start of the collapse, is discussed in HW13.

\paragraph{Carbon, nitrogen, and oxygen reservoirs}
The main carbon-bearing molecules of MU10$\Theta$0 are, from the most to the least abundant, CH$_4$, CO, H$_2$CO, CH$_3$OH, and CH$_3$C$_2$H.
Their mean total abundances are respectively $5.6\times10^{-5}$, $5.5\times10^{-5}$, $2.0\times10^{-5}$, $9.0\times10^{-6}$, and $5.9\times10^{-6}$.
The values do not depend significantly on the component.
These carbon-bearing molecules are mainly located on the grain surface due to the density and temperature conditions, except in the core where they are in the gas phase.
These results are similar for all models. 
As an example, the total abundance of CO in the envelope and the outflow of MU2$\Theta$0 is $5.4\times10^{-5}$, and the total abundance of CH$_3$OH is $9.3\times10^{-6}$. These values  are very close to the values obtained with the MU10$\Theta$0 core model.

The main nitrogen-bearing molecules of MU10$\Theta$0 are, from the most to the least abundant, NH$_3$, N$_2$, HCN, CH$_3$NH$_2$, and HNC.
Note that the main nitrogen reservoir is NH$_3$ instead of N$_2$, due to the values of some gas-phase rate constants that we have taken into account following \cite{daranlot_elemental_2012}.
The mean total abundances of NH$_3$, N$_2$, and CH$_3$NH$_2$ are respectively $3.1\times10^{-5}$, $1.2\times10^{-5}$, and $2.3\times10^{-6}$ in all models and components.
Contrary to HCN, which has a total abundance of about $3-4\times 10^{-6}$ in all models and components, the HNC total abundance depends more on models and components than the other main nitrogen-bearing molecules, and therefore would be a good chemical tracer {of the intensity and orientation of the magnetic field and the nature of the component}.
Its total abundance is $(4.6-4.7)\times10^{-7}$ in the envelope and the pseudodisk of every model, except in the pseudodisk of MU200$\Theta$0 where its value decreases to $5.9\times10^{-8}$.
In the disk of MU10$\Theta$45, its total abundance is $1.4\times10^{-7}$, and it is 3 times higher for MU10$\Theta$0, whereas it is 30 times lower for MU200$\Theta$0.
Its total abundance range is $(3.6-4.7)\times 10^{-7}$ in the outflows of the different models, and is $(1.3-2.5)\times 10^{-8}$ in the central cores.
HNC is very dependent on the temperature conditions.
After its desorption from the grain, it can be quickly destroyed in the gas phase through ion neutral reactions due to its highly polar nature, and as a consequence its total abundance can be highly decreased \citep[see also][]{hincelin_survival_2013}.
The ion-neutral chemistry involved in the destruction of HNC has a timescale of about 50~yr at 100~K and $10^{10}$~cm$^{-3}$, much lower than the dynamical timescale of the disk of the MU200$\Theta$0 model which is about 1,200~yr (see also Section~\ref{subsec:timescales}).
{Results on gas phase HNC and HCN will however need to be confirmed using more recent data on their chemistry \citep{loison_interstellar_2014}.}

The main oxygen-bearing molecules are H$_2$O, CO, H$_2$CO, and CH$_3$OH.
The mean total abundance of water relative to the total density of hydrogen is $6.5\times 10^{-5}$, and it depends neither on component nor on model.
Water is principally located on the grain surface except in the warmer portion of the central core.
Other oxygen-bearing species, such as O, O$_2$ and CO$_2$, are not very abundant due to the oxygen-poor elemental abundances we have chosen to reproduce observations of molecular clouds \cite[see][for a detailed discussion]{hincelin_oxygen_2011}.
Abundances of these three species are $10^{-9}$ to $10^{-17}$, $10^{-11}$ to $10^{-13}$, and about $3\times 10^{-8}$ respectively, depending on the component.
The high density also increases the depletion of gas-phase atomic and molecular oxygen which then react with H on the grain surface to form water ice \citep{hincelin_oxygen_2011}.

\paragraph{Charge carriers}
The dominant charged species in the envelope do not depend on models, and are the following: electrons, H$_3^+$, N$_2$H$^+$, and H$^+$, with abundances relative to the total density of hydrogen ranging from $\sim10^{-9}$ to $\sim10^{-10}$.
This is also true in pseudodisks of all models, except MU200$\Theta$0 where negatively charged grains become the second most abundant charge carrier.
Since MU200$\Theta$0 presents fragmentation, the pseudodisk includes high density regions surrounding the fragments.
Thus, {recombinations between electrons and cations in the gas phase are more frequent}, which explains the larger {relative} abundance of these charged grains {compared to other charged species}.
From a general point of view, the denser the medium, the higher the {relative} abundance of negatively charged grains, {and the lower the ionization fraction \citep{nakano_mechanism_2002}.}
This result is directly linked to the estimate of non-ideal MHD resistivities \citep[Ohmic, ambipolar and Hall, e.g.,][]{li_non-ideal_2011} which are computed from the abundances of the different charge carriers.
The resistivities regulate the transport of magnetic flux and of angular momentum in collapsing cores, as found in recent non-ideal MHD simulations \citep[e.g.,][]{li_earliest_2014,tomida_radiation_2015,tsukamoto_effects_2015}.
This result is particularly highlighted in disks and cores {where charged grains} become the second or the first most abundant charge carrier.
{For example in the MU10$\Theta$0 model, the ionization fraction is $4.6\times 10^{-12}$ and $9.0\times 10^{-12}$ respectively in the core and the disk, while it is $4.8\times 10^{-9}$ in the envelope.
The contributions of negatively charged grains to the ionization fraction are 39~\%, 20~\%, and 0.04~\% respectively in the core, the disk, and the envelope.}

\paragraph{Complex organic molecules}
We consider as Complex Organics Molecules (COMs), molecules with six atoms or more, and which contain the element carbon \citep{herbst_complex_2009}.
The most abundant COM produced in our simulations is methanol (CH$_3$OH), which is located on the grain surface (except in the central core, where it has been desorbed).
Its mean total abundance of $9.0 (\pm0.1)\times10^{-6}$ is roughly the same in all models and all components.
Once CO is formed in the gas phase and is adsorbed on grain surfaces, it is relatively efficiently hydrogenated, due to diffusion of hydrogen atoms following the Langmuir-Hinshelwood mechanism, into HCO, H$_2$CO, H$_2$COH, and finally CH$_3$OH \citep{charnley_deuterated_1997}.

The most abundant oxygen and nitrogen bearing COMs in envelopes, after methanol, are CH$_3$NH$_2$ (with a mean total abundance of $2\times10^{-6}$), C$_2$H$_5$CN ($3\times10^{-7}$), H$_2$C$_3$O ($2\times10^{-8}$), CH$_3$CN ($9\times10^{-9}$), and NH$_2$CHO ($8\times10^{-9}$).
These COMs are formed at low temperature and high density ($\sim10$~K and $\geq10^5$~cm$^{-3}$) on the grain surface by successive H atom recombinations \citep{hasegawa_models_1992}.
The chemistry is already active during the cloud phase, before the collapse, towards the end of which their abundances are already high.
For example, the formation of methylamine (CH$_3$NH$_2$) starts from the recombination on the grain surface between N and CH$_2$.
Then, the product H$_2$CN is successively hydrogenated to form CH$_3$N, CH$_2$NH$_2$, and finally CH$_3$NH$_2$.
The formation of acetonitrile (CH$_3$CN) and cyclopropenone (H$_2$C$_3$O) start from the surface reaction $\rm N + C_2$ and $\rm O + C_3$, respectively, and then H atoms recombine with the products.
These COMs stay on the grain surface in the ices, until they come close to the central core itself or the inner region of the disk.
However, COMs such as $\rm CH_3OCH_3$ and HCOOCH$_3$ are not abundant during this period, with a low total abundance of around $1 \times 10^{-13}$ and $4 \times 10^{-13}$ respectively at the end of the cloud phase, and not much more in the cold collapsing envelope.
These molecules seem to form on the surface of grain only by recombination of radicals, as explained by \cite{garrod_complex_2008},
and so they need a higher temperature to allow some diffusion.
Recent observations however show a non-negligible gas-phase abundance of $\sim 10^{-11}$ or more for CH$_3$CHO, CH$_3$OCH$_3$, HCOOCH$_3$, and other COMs in cold prestellar cores \citep{bacmann_detection_2012,vastel_origin_2014} and cold envelopes of low-mass protostars \citep{oberg_cold_2010,jaber_census_2014}, which challenge existing chemical models.
The origin of these COMs is not very well understood yet, but some explanations have been proposed involving e.g. non-thermal desorption \citep{vasyunin_reactive_2013,chang_interstellar_2014,vastel_origin_2014}, Eley-Rideal and complex inducing reactions \citep{ruaud_modelling_2015}, and gas-phase formation routes involving halogen atoms and radiative association \citep{balucani_formation_2015,vasyunin_reactive_2013}.
Envelopes also contain many hydrocarbons which are relatively abundant compared to the other abundant COMs (CH$_3$OH, CH$_3$NH$_2$, etc); e.g., CH$_3$C$_2$H ($6\times10^{-6}$), C$_2$H$_6$ ($2\times10^{-6}$), C$_4$H$_4$ ($2\times10^{-7}$), C$_5$H$_4$ ($7\times10^{-8}$), $\rm C_9H_4$, C$_7$H$_4$, and C$_6$H$_4$ ($1\times10^{-8}$), and $\rm C_8H_4$ ($3\times10^{-9}$).
These long chain molecules are formed due to the carbon-rich elemental abundances ($\rm C/O>1$; \cite{wakelam_effect_2006,hincelin_oxygen_2011}).

Among COMs cited above, the nature and mean total abundance of the molecules in pseudodisks, outflows, and disks are roughly the same as those in envelopes, except for the MU200$\Theta$0 model.
The pseudodisk of model MU200$\Theta$0 holds a slightly higher total abundance of NH$_2$CHO ($1\times10^{-8}$), while a significant higher total abundance of this molecule is seen in the disk ($6\times10^{-8}$).
When H$_2$CO, which has been formed on the grain surface during the cold phase of the cloud, desorbs from the grain surface during the collapse at around 40~K, a larger total abundance of this molecule in the gas phase enhances the formation of NH$_2$CHO through the gas phase reaction $\rm NH_2 + H_2CO$\footnote{A new value of the rate coefficient has been calculated by \cite{barone_gas-phase_2015}, which was not available at the time of our simulation. The reaction is found to be almost barrierless,  which facilitates the gas-phase formation of formamide at low temperature, without the need of grain-surface chemistry if NH$_2$ and H$_2$CO are abundant enough in the gas phase. We included this reaction at the time of our calculation, with an estimated rate coefficient of $10^{-10}$~cm$^3$s$^{-1}$ according to the KIDA database and with the assumption that the reaction is barrierless.}.
In the disk of the MU200$\Theta$0 model, the mean abundance of HCOOCH$_3$ is  enhanced due to grain surface reactions to a value equal to $2\times10^{-9}$.
Carbon chains, such as $\rm C_2H_4$, $\rm C_6H_6$, $\rm C_8H_4$, and $\rm C_6H_2$ are also formed more efficiently than in other models, because of the medium temperature and high density that respectively promotes grain-surface diffusion and increases adsorption.

Due to a significantly higher mean temperature, cores present higher total abundances for some COMs.
NH$_2$CHO is more abundant in all cores, compared to the other components, by roughly a factor of 10.
HCOOCH$_3$ is also very abundant in cores compared to other components, reaching a mean value of $10^{-8}$ in MU10$\Theta$45 and $(2-4)\times10^{-9}$ in other models.
Long carbon chains are enhanced in cores to values in the range $(1-10)\times10^{-9}$.

\begin{figure*}
\centering
\includegraphics[width=1.0\linewidth]{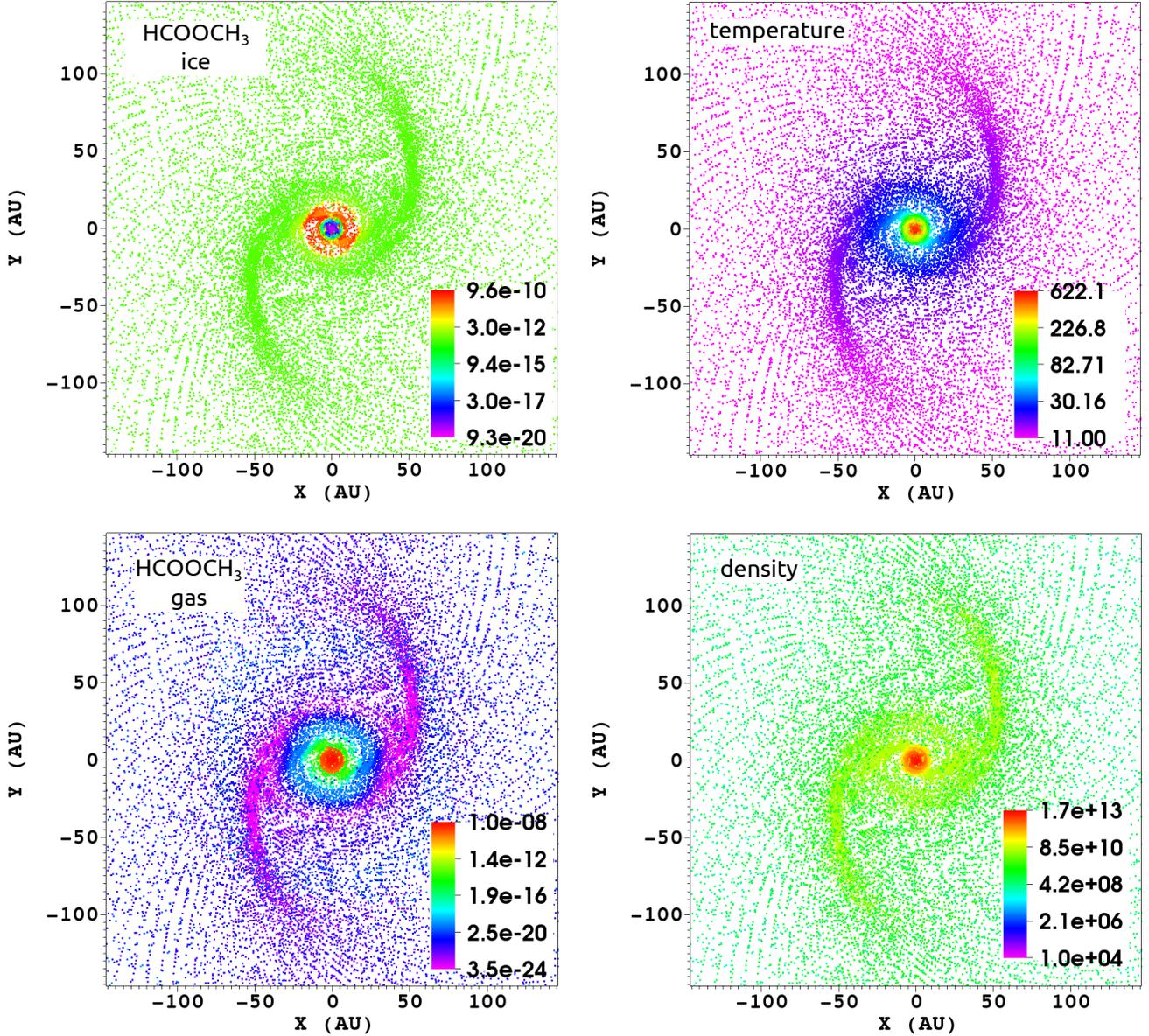}
\caption{Methyl formate (HCOOCH$_3$) abundances on the grain surface (top left) and in the gas phase (bottom left) relative to total hydrogen nuclei, for particles of the MU10$\Theta$0 model, at the end of the simulation.
Particles belong to a $(\rm 100~AU)^3$ cube centered on the first Larson core, and are projected onto the x-y plane.
Corresponding temperature (K) and density (cm$^{-3}$) maps are given in the right panel.}
\label{fig:HCOOCH3_mu10theta0}
\end{figure*}

Note that these values are averaged over the total number of particles that belong to a component, so they do not reflect the variety of abundances that can be present within one component.
A good example of "active chemistry" (as opposed to physical mechanisms such as desorption and adsorption) inside a specific area of the collapsing dense core is given by methyl formate (HCOOCH$_3$) as shown in Figure~\ref{fig:HCOOCH3_mu10theta0}.
This figure shows the abundance of methyl formate in the MU10$\Theta$0 model, in the gas phase and on the grain surface, including every particle inside a $100^3$~AU$^3$ cube projected on the x-y plane.
Methyl formate is efficiently formed on the grain surface, in a ring of matter around the central core, from 8 to 12~AU from the center.
Its abundance in the ices is enhanced from $\sim10^{-13}$ outside the ring to $\sim10^{-9}$ in the ring.
{This region follows the chemistry of COM formation of \cite{garrod_complex_2008} but at an early stage.
The high densities involved enhance chemical rates and decrease chemical timescales.}
{Inside the inner 8~AU, methyl formate is desorbed and thus its gas-phase abundance is strongly enhanced.}
However, we must show caution about its abundances in the very center of the central core, where the temperature is larger than 300~K, since our chemical network does not take into account accurate rate constants for chemistry above 300~K.
We will improve this point in future simulations following \cite{harada_new_2010,harada_erratum:_2012}.

\subsubsection{Chemical evolution along trajectories during the collapse}

\begin{figure}
\centering
\includegraphics[width=1.0\linewidth]{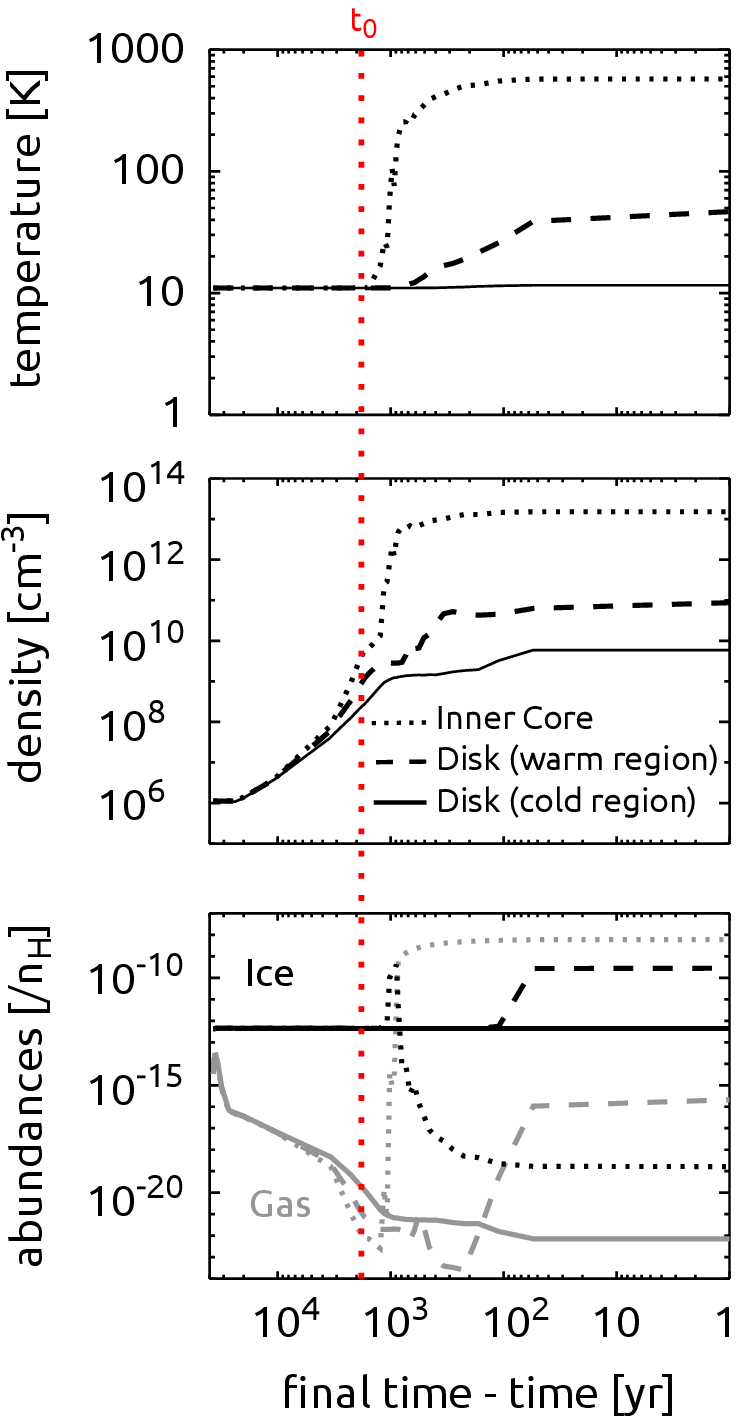}
\caption{Evolution of the temperature (top panel) and the density (middle panel) of three different particles that belong to the MU10$\Theta$0 model.
The particles end their trajectory in the central core (dotted line) and the cold (solid line) and the warm (dashed line) regions of the disk.
The bottom panel shows the associated abundances of methyl formate HCOOCH$_3$ in the gas phase (gray) and the ice (black).}
\label{fig:HCOOCH3_evolution}
\end{figure}

As mentioned in the previous section, although adsorption and desorption dominate, chemistry may have time to produce molecules even if the collapsing time is short.
This is due to the very high density, close to $10^{10}$~cm$^{-3}$ or even higher depending on the region, which greatly enhances the total rates of chemical reactions.
As an example, Figure~\ref{fig:HCOOCH3_evolution} gives the evolution of the abundance of HCOOCH$_3$ as a function of time, along the trajectories of three different particles that belong to the MU10$\Theta$0 model.
One particle ends up inside the central core, while the two others end up in a cold and in a warm region inside the disk.
The figure also displays the temperature and the density of the particles as a function of time.
The evolution can be divided up to three different phases depending on the particle.
During the first one, the molecules are mainly present in the ice due to the low temperature, and the gas phase molecules are more and more depleted as the density increases with time.
It occurs for all three trajectories until about 2000~yr before the final time.
During the second phase, the temperature rises, and methyl formate is efficiently formed on the grain surface through the radical-radical reaction HCO + CH$_2$OH.
It occurs for both particles which end in the core and in the warm region of the disk, but at different times, 1000 and 100~yrs before the final time respectively.
Then, if the temperature is high enough, methyl formate desorbs.
This last phase occurs only for the particle which reaches the central core, 800~yrs before the end.
Other complex organic molecules, such as CH$_3$C$_2$H, CH$_3$OCH$_3$, and HC$_7$N, present a similar behavior.
The abundance of NH$_2$CHO also increases along the trajectories of the tracer particles, but it is formed in the gas phase.
Its abundance is increased because of desorption of reactants involved in its gas-phase formation, as described in the previous section.

\begin{figure*}
\centering
\includegraphics[width=1.0\linewidth]{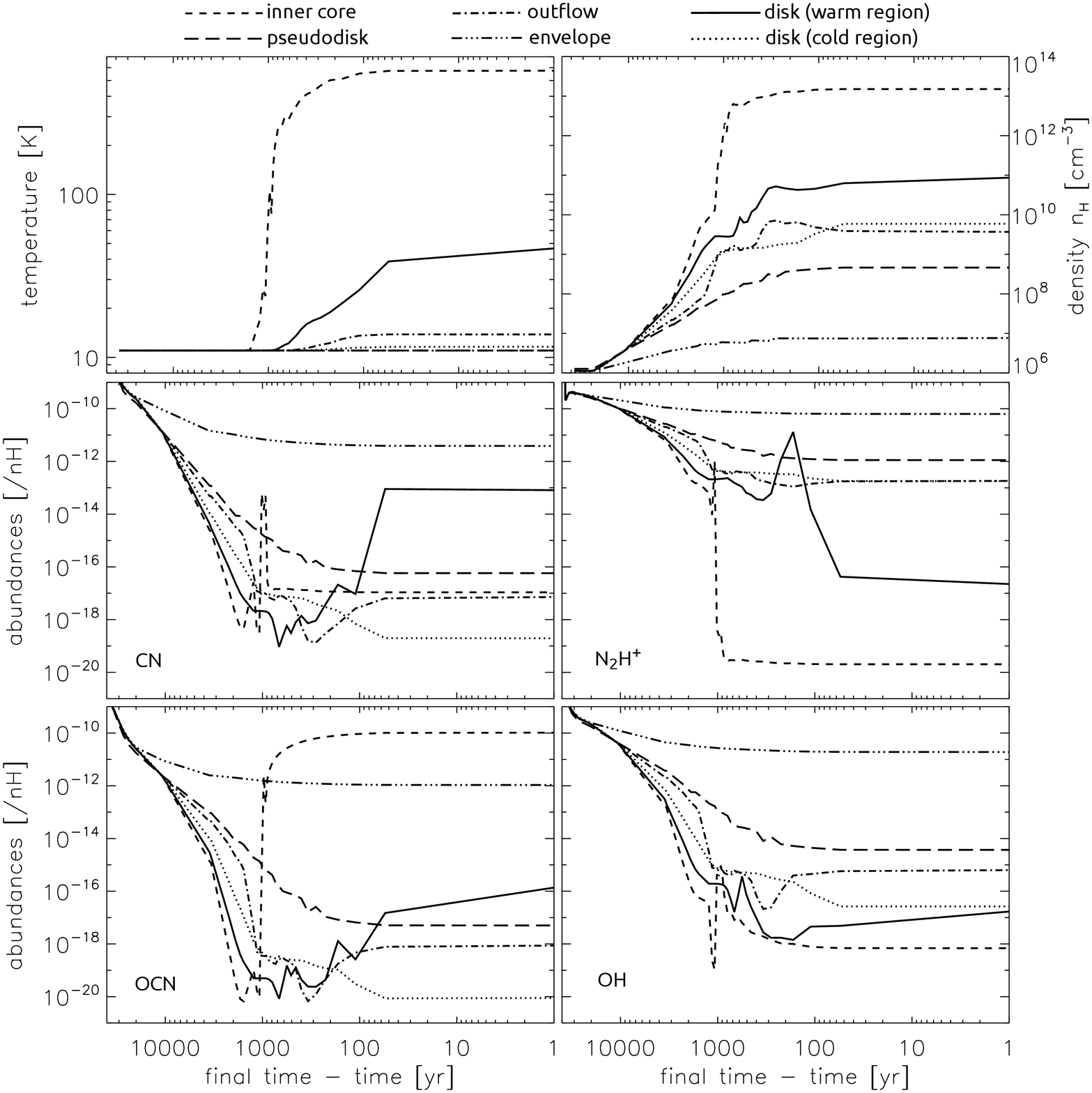}
\caption{Temperature and density as a function of time along six different trajectories, and the corresponding 
gas phase abundances of CN, OCN, N$_2$H$^+$, and OH.
The tracer particles end their trajectories inside different regions of the MU10$\Theta$0 model: the central core (dashed line), the pseudodisk (long dashed line), the outflow (dashed and single dotted line), the envelope (dashed and triply dotted line), the warm region of the disk (solid line), and the cold region of the disk (dotted line).
The particles that end in the central core and the disk are the same ones as in Figure~\ref{fig:HCOOCH3_evolution}.}
\label{fig:evol_component}
\end{figure*}

As seen in Figure~\ref{fig:HCOOCH3_evolution} for HCOOCH$_3$, the abundances may vary depending on the component of a given model.
Figure~\ref{fig:evol_component} presents the abundances as a function of time for some particularly sensitive species: CN, OCN, OH, and N$_2$H$^+$.
The abundances evolve along different trajectories of the MU10$\Theta$0 model, i.e. for different tracer particles that end inside the central core, two different regions of the disk, the pseudodisk, the outflow, and the envelope.
For a given species, the chemical evolution can be highly dependent on the trajectory.
For example, while the gas phase abundance of CN in the envelope is a few $10^{-12}$ at the final time, its abundance is about $10^{-17}$ in the outflow and the central core.
On the other hand, very different physical conditions between the outflow and the central core can result surprisingly in roughly the same abundance of CN.
While in the central core, CN is first desorbed and then destroyed by gas phase neutral-neutral reactions, in the outflow, CN is first adsorbed and then regenerated by gas phase reactions between electrons and ions.
Besides the dependence on the trajectory, we also observe a dependence on the species for one given trajectory.
While the abundances are all decreasing in the envelope for the four species -- due to adsorption of the species or their precursors because of the increasing density and the low temperature -- they can differ for other components.
On the one hand, the abundances of CN, OCN, and OH for the warm disk all tend to decrease until a few hundred years before the final time, and then rise.
On the other hand, we observe a significant increase of the abundance of N$_2$H$^+$, about two orders of magnitude, around 200~yr before the final time, and then a huge decrease, about six orders of magnitude.
N$_2$H$^+$ is first efficiently formed by the reaction $\rm H_3^+ + N_2$.
Then, CO and CH$_4$ desorb and quickly destroy N$_2$H$^+$.

\begin{figure*}
\centering
\includegraphics[width=1.0\linewidth]{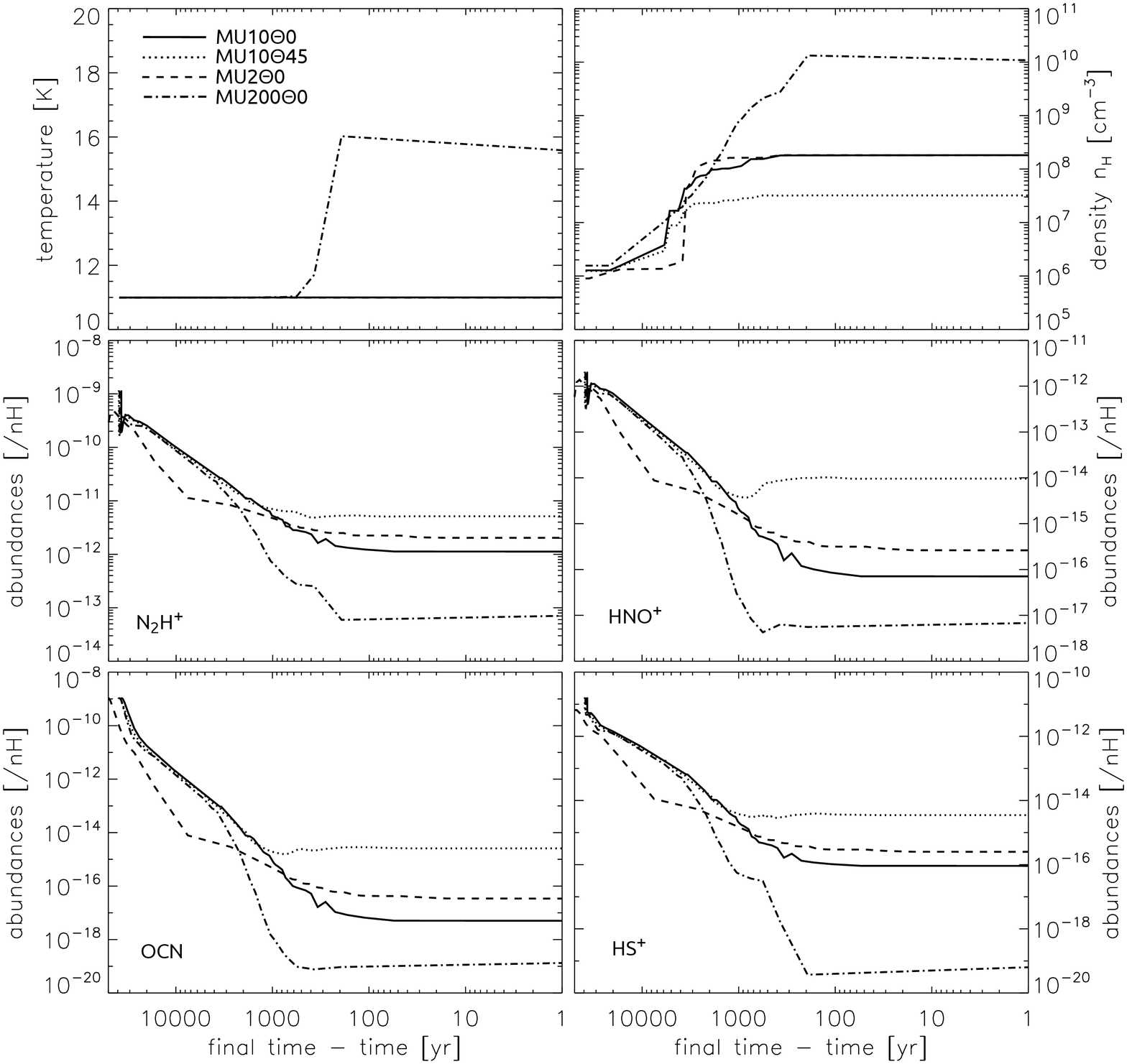}
\caption{Temperature and density as a function of time along four different trajectories, and the corresponding gas phase abundances of N$_2$H$^+$, HNO$^+$, OCN, and HS$^+$.
The tracer particles end their trajectories inside the pseudodisk of the four models: MU10$\Theta$0 (solid line), MU10$\Theta$45 (dotted line), MU2$\Theta$0 (dashed line), and MU200$\Theta$0 (dashed and single dotted line).
The particle that ends in the pseudodisk of MU10$\Theta$0 is the same one as in Figure~\ref{fig:evol_component}.}
\label{fig:evol_models_pdisk}
\end{figure*}

Abundances may also be sensitive to the chosen model, for the same component, because the matter may have a different physical history.
To give an example, Figure~\ref{fig:evol_models_pdisk} presents the chemical evolution of N$_2$H$^+$, HNO$^+$, OCN, and HS$^+$ for four different particles (one for each model) that end their trajectory inside the pseudodisk.
We do not observe a significant difference among the abundances of one given species for the four trajectories, until 2000~yr before the end of the simulation.
After this time, the abundances tend to split more or less widely, from a factor of a few to several orders of magnitude.
These different abundances obviously come from the different physical conditions the particles undergo during the collapse.
Note that the present physical conditions may not reset the effect of the past history.
This is the case for the two particles of MU10$\Theta$0 and MU2$\Theta$0.
Their temperature evolutions are the same, but their density evolutions differ.
The particle from the second model undergoes a steeper variation.
Even though its initial and final densities are the same as the other particle, the abundances of the four species are still finally different by a factor 2 to 6 depending on the species.
Even if these factors are small compared to the typical uncertainty of a gas-grain code -- about one order of magnitude in abundance -- this result yields important information on a possible degeneracy.
The same physical condition at a given time may correspond to different chemical results, because a fraction of the past dynamical history was different from one component to the other.
This difference induces some possible limitation on the precise derivation of the dynamical history from observed abundances.

\subsection{Chemical versus dynamical timescales}
\label{subsec:timescales}

A comparison between chemical timescales and  dynamical timescales is helpful to verify the usefulness of the time dependency of both physics and chemistry in parallel.
If these timescales are similar, then pseudo-time dependent models, where physics or chemistry is evolving while the other is fixed, may be less appropriate to compute an accurate chemical evolution.

On the one hand, the dynamical timescale is specific to the considered component.
The dynamical timescale of the collapsing envelope can be computed considering a free-fall collapse, and
is equal to $4.5\times 10^4$~yr.
The dynamical timescale of the disk is given by its rotational period.
According to the azimuthal velocities of the matter, this period is about 1,500~yr. 
The dynamical timescale of the outflow is given by its radial velocity:
500 to 1,000~yr is the time required by the matter to go from the central core to the upper edge of the outflow, depending on the core model.
Finally, the dynamical timescale of the central core can be computed using the sound speed, which is about 20~yr.

On the other hand, the chemical timescale is not a unique quantity.
For our problem, desorption and adsorption timescales are involved, as well as gas-phase and grain-surface reaction timescales.
The chemical timescale is often dependent on the temperature, the density, the considered species and their abundances, and the time, so it is not a simple quantity to estimate and needs to be used with caution.
We refer to Appendix~\ref{appendix:timescales} for a detailed calculation of adsorption and desorption timescales and to Table~\ref{tab:ads_des_timescales} for an application of the calculation for CO.
While adsorption timescales are similar to or lower than the different dynamical timescales, desorption  occurs as quickly as the dynamics once a temperature threshold is surpassed, which is  20~K for CO for example.
As examples, the adsorption timescale for CO is about $4\times 10^3$~yr in the collapsing envelope, which is about 10~\% of its dynamical timescale.
However, the desorption timescale for CO can be considered to be infinite (about $10^{20}$ times the age of the universe) and thus is not efficient in the envelope.
Ion-neutral reactions form the bulk of the gas-phase chemistry and their timescale is computed by
\begin{equation}
\tau_{i-n} = \left( k_{i-n} [ions] \right)^{-1} = \left( k_{i-n} f_i n \right)^{-1},
\end{equation}
where $k_{i-n}$ is the rate coefficient for ion-neutral reactions (in cm$^3$s$^{-1}$), $[ions]$ is the concentration of ions (in cm$^{-3}$), $f_i$ is the ionization fraction, and $n$ is the gas-phase density (in cm$^{-3}$).
The rate coefficient $k_{i-n}$ is generally about $10^{-9}$~cm$^3$s$^{-1}$, the so-called Langevin value, which leads to a
chemical timescale of approximately $3\times 10^4$~yr in the envelope, 300~yr in the disk, and 1~yr in the central core.
It is similar to the dynamical timescale of the envelope, while it is about a factor of 10 smaller than the dynamical timescales of the disk and the core, which means that gas-phase chemistry influences  the system much faster in the disk and the core than in the envelope during the dynamical evolution of the matter.

The grain-surface chemical timescale is given by  
\begin{equation}
\tau_{surf} \simeq \frac{1}{k_{diff} n_{ice}} \simeq \frac{n_d}{R_{diff} n_{ice}} = \frac{n_dN_s}{\nu \exp(-\frac{E_b}{k_BT}) n_{ice}},
\end{equation}
where $n_{ice}$ is the concentration (in cm$^{-3}$) of ice, $n_d$ the granular concentration (in cm$^{-3}$), $N_s$ the total number of sites on the surface of one grain, $\nu$ the characteristic adsorbate vibrational frequency, $E_b$ the energy barrier against diffusion, and $R_{diff}$ the thermal diffusion rate.
If we compute the timescale for CO to react on the surface -- assuming it can react with the entire ice,  which means that this timescale is a lower limit -- with a fractional abundance of ice equal to $10^{-4}$, we obtain $2\times 10^4$~yr at 10~K, and less than a year at 15 and 20~K.
We however emphasize that the grain-surface chemistry timescale is meaningful once molecules are adsorbed on the grain surface, so it has to be added to several adsorption timescales to be compared with the dynamical timescale.
We see that in the envelope, the grain-surface chemistry does not have time to efficiently change the chemical composition.
Because of the long timescale of this process in the envelope, the chemical steady-state is shifted to longer times compared with pure gas-phase chemical modeling.
Note that chemical equilibrium is not reached either because reverse reactions do not occur in our system, mainly prevented by their endothermicity.  See Appendix~\ref{appendix:ss_vs_eq} for more detail on chemical steady-state and equilibrium.

Several gas-phase chemical timescales are needed to significantly modify the abundance of species.
The resulting timescale is longer than the time for the initial uniform sphere to change to a Bonnor-Ebert-like density profile, which is less than $10^4$~yr in our simulation.
Adsorption starts to occur during this transitional period, but the grain-surface chemical timescale is very long at 10~K.
Abundances of species are thus not significantly dependent on our initial physical structure.

To conclude, the chemical timescale is often not negligible compared to the dynamical timescale and vice-versa, which shows the importance of modeling both chemical and dynamical evolution in parallel.

\section{A chemical distinction among components and core models}
\label{subsec:tracers}

Some chemical species may only exist in a few components or a few core models, or at least present a huge variation of their abundance among these different regions.
Thus, they could be good candidate as tracers.
We try to identify such species and present the results in this section.
Our aim is also to identify tracers among these species that are specific to the components regardless of the core model, and vice-versa.
We refer to the previous sections for more detail about the physical processes and the chemical reactions responsible for the variation of the abundances.
{Our identification is only based on the predicted abundances.
Detectability of the species using 3D radiative transfer modeling -- such as RADMC-3D \citep{dullemond_radmc-3d:_2012} -- needs a dedicated study and will be the subject of future work.}

{We have used the following formula to identify the gas-phase species $s$ with the largest variation of mean abundances\footnote{Contrary to section~\ref{subsubsec:distri_molecules}, gas phase and grain surfaces abundances are not added together, and the calculation is done for gas-phase species.} $<A>$ between two components of the same model or between two models and the same component:}
\begin{equation}
\Sigma(s) = \sum\limits_{i\neq j} |<A(s)>_i - <A(s)>_j|
\end{equation}
where $i$ and $j$ are the components of a given model (first case), or the models of a given component (second case).
The higher $\Sigma(s)$, the more sensitive the species $s$.
Tables~\ref{tab:chem_trac_components} and \ref{tab:chem_trac_models} present identified gas-phase chemical species that are particularly sensitive, {along with species observed in first Larson core candidates (see Section~\ref{sec:obs}) and protoplanetary disks}.
Table~\ref{tab:chem_trac_components} facilitates the comparison of the different components for a given model, while Table~\ref{tab:chem_trac_models} facilitates the comparison of the different models for a given component.
We have selected only species with an abundance larger than $10^{-12}$ in at least one component or one model.
Abundances of species highlighted in red are higher than those in other components or models, while species highlighted in blue are depleted.
The symbol * highlights promising tracers whose abundances are at least higher by a factor 10 in one unique component for a given model, or one unique model for a given component.
These tracers are particularly promising to identify the core model or the component.
We discuss them in the summaries of Sections~\ref{subsec:tracers_comp} and \ref{subsubsec:tracers_cores}.

Tracers presented in this section may not be exclusive to the first Larson core, and may also pertain to the second Larson core.
However, \cite{vaytet_simulations_2013} show a factor of 2 or more difference in the radial profiles of the temperature of the first core compared to the second core, depending on the radius.
This difference is non negligible for the chemistry, thus tracers may change for the second Larson core.
A detailed simulation of the gas-grain chemistry during the formation of the second Larson core will be necessary to confirm this hypothesis.

\subsection{Differences among components for a given model}
\label{subsec:tracers_comp}

{
{As a general trend}, species sensitive to components are more abundant in the envelope or the central core, while they are less abundant in the outflow, the disk, and the pseudodisk.
The outflow, the disk, and the pseudodisk are in general cold and dense, which promotes adsorption and explains the depletion of gas-phase species, while the high temperature of the central core enhances desorption and gas-phase reactions involving the newly desorbed species (see Section~\ref{subsec:chem_charac}).
The species CN, HCO$^+$, N$_2$H$^+$, HNO, C$_2$H, and H$_2$CN trace the envelope, while the other species of Table~\ref{tab:chem_trac_components} trace the central core.
N$_2$H$^+$ for example is formed in the cold envelope in the gas-phase, but at higher temperature -- in the central core -- once CO and CH$_4$ desorb, it is destroyed in the gas-phase by these two species (see Section~\ref{subsec:chem_charac}).
A majority of species shows important difference between the envelope and the outflow or the pseudodisk, allowing a chemical distinction between the envelope and these two other components.
We present the results for each model in detail.}

\paragraph{$\rm {MU2\Theta0}$}

{
CN, HCO$^+$, N$_2$H$^+$, C$_2$H, and H$_2$CN are more abundant by at least two orders of magnitude in the envelope than in any other component, so are very specific to this component.
The molecules CO, HCN, HNC, H$_2$O, HNCO, H$_2$CO, NH$_3$, HC$_3$N, C$_3$H$_2$, and CH$_3$OH show high abundances in the central core, so are specific to this component.
All the species presented in the Table~\ref{tab:chem_trac_components} but CN, HCO$^+$, and N$_2$H$^+$ are mostly depleted in the pseudodisk.
Thus, observing a depletion of these species in a disk shaped region within a dense core could trace the pseudodisk of a highly magnetized core.
}

\paragraph{$\rm {MU200\Theta0}$}

{
The same group of species as in the MU2$\Theta$0 model (CN, HCO$^+$...) plus HNO are specific to the envelope.
The other same group of species (CO, HCN...) are also specific to the central core.
Only the three species HCN, H$_2$O, and NH$_3$ however are depleted in the pseudodisk.
For some species, such as CS, HNC, H$_2$O, H$_2$CO, and NH$_3$, each component presents a different abundance from the others, which makes these species particularly interesting to trace the components.
For the disk and the central core, the ratio HCN to HNC is very far from unity (around 1000).Gas-phase ion-neutral reactions consume HNC more efficiently than HCN in these warmer components (see Section~\ref{subsec:chem_charac}).
}

\paragraph{$\rm {MU10\Theta0}$}

{
CN, HCO$^+$, N$_2$H$^+$, C$_2$H, H$_2$CN, and HNO are specific tracers of the envelope.
As highlighted in red in Table~\ref{tab:chem_trac_components}, all the abundant species of the central core (CO, CS, HCN, HNC...) are more abundant by at least one order of magnitude in this component than in any other component.
Contrary to the two previous models, the disk is the component where species are mostly depleted.
The disk of this model is dense and on average colder than the disk of the other models, which favors adsorption.
The species CO, N$_2$H$^+$, HNO, H$_2$CO, and NH$_3$ allow differentiation between the disk and the pseudodisk.
The outflow is distinguishable from the envelope using CO, N$_2$H$^+$, HNO, H$_2$CO, and NH$_3$.
}

\paragraph{$\rm {MU10\Theta45}$}

{
CN, N$_2$H$^+$, HNO, and C$_2$H are more abundant by several orders of magnitude in the envelope than in any other component.
HCO$^+$ and H$_2$CN are also abundant in the envelope, but a similar value is obtained in the outflow for the first species, and in the central core for the second one.
As in the MU10$\Theta$0 model, all the abundant species of the central core (CO, CS, HCN, HNC...) are more abundant by at least one order of magnitude in this component than in any other component.
The disk and the pseudodisk are the two regions where depletion of gas-phase molecules is the most important.
The species CO, N$_2$H$^+$, and NH$_3$ allow differentiation between the disk and the pseudodisk.
The outflow is distinguishable from the envelope using CO, N$_2$H$^+$, HNO, H$_2$CO, and NH$_3$.
}

\paragraph{Summary}
Some tracers are specific to the components regardless of the core model, which make them very useful to identify the components of a first Larson core if we do not know the properties of the core (magnetization level and inclination of the rotational axes).
The envelope can be identified by a high abundance of CN, HCO$^+$, N$_2$H$^+$, HNO, C$_2$H, and H$_2$CN.
The central core is traced by CO, CS, HCN, HNC, H$_2$O, HNCO, H$_2$CO, NH$_3$, HC$_3$N, C$_3$H$_2$, and CH$_3$OH.
In the other components, the abundances of the species are lower than in the central core or lower than in the envelope, which makes the identification of the components more complicated.
However, if a disk-like structure is observed, a distinction can be made between a disk and a pseudodisk regardless of the core model.
The HCN to HNC ratio is higher than 30 in the disk compared to the pseudodisk in every core model, and H$_2$O, HCO$^+$, and N$_2$H$^+$ are more abundant by an order of magnitude or more in the pseudodisk than in the disk, except for water for the MU200$\Theta$0 core model because of its desorption from the grain surface due to the presence of the hot fragments within the disk.
With a high enough resolution, observation of fragmentation can however discriminate this last core model.
Finally, our results show that CS, CN, N$_2$H$^+$, C$_2$H, and H$_2$CN are less abundant by at least two orders of magnitude in the outflow than the envelope regardless of the core model -- mainly due to adsorption of the species or their precursors -- which helps to discriminate the outflow from the envelope.

\subsection{Differences among core models for a given component}
\label{subsubsec:tracers_cores}

{Using Table~\ref{tab:chem_trac_models}, which displays the most sensitive species to a change in the model for a given component, we extract the most promising species to trace the magnetization level and the $\Theta$ angle of a core model.}

\paragraph{Envelope}

The abundance of a given species within the envelope varies by at most a factor of two among the different models, so the collapsing envelope does not seem to be a good target to distinguish one model from the other.
This result was expected, since the mean history of the matter within the envelope does not vary much from one model to the other (see section~\ref{subsubsec:phys_history_matter}).
The result is however different for the following other components.

\paragraph{Central Core}

On the one hand, HNO is much more abundant in the central core of the MU2$\Theta$0 model than the others, thus is a potential good tracer of a highly magnetized core.
On the other hand, H$_2$CN is much more abundant in the central core of MU10$\Theta$45 model than the other models, thus is a potential good tracer of cores with a misalignment between the magnetic field and the rotational axis.
{The central core of MU200$\Theta$0 is characterized by species with a lower abundance (highlighted in blue in Table~\ref{tab:chem_trac_models}).}

\paragraph{Outflow}

The outflow of the MU2$\Theta$0 model is well distinguished by the enhanced abundance of a majority of species, which are greater by several orders of magnitude compared to other models, {while the outflow of the MU10$\Theta$0 model shows depletion of species.}
The outflow of the MU2$\Theta$0 model is on average 3 to 4 times warmer than the other which explains the observed differences.
The abundances for the MU10$\Theta$45 model are generally higher than for the MU10$\Theta$0 model, {by a factor of a few to one or two orders of magnitude depending on the species.}
{HCO$^+$ and N$_2$H$^+$ are especially abundant in the MU10$\Theta$45 model.
For this model, the temperature is sufficiently low that CO does not desorb much, which prevents the destruction of N$_2$H$^+$, and the average density is lower than in the other models, which decreases the recombination between electrons and these two cations.}

\paragraph{Disk}

In general, abundances of gas-phase species are increased in the sequence going from the disk of MU10$\Theta$0 {to MU10$\Theta$45  to MU200$\Theta$0.}
{As an example, CO is more abundant in the disk of MU200$\Theta$0 than in the disks of MU10$\Theta$45 and MU10$\Theta$0 by about one and three orders of magnitude, respectively.}
{CO and HNO allow a good distinction between the MU10$\Theta$0 and MU10$\Theta$45 models.}

\paragraph{Pseudodisk}

H$_2$CO trace the MU200$\Theta$0 model, while N$_2$H$^+$ is significantly depleted in the MU200$\Theta$0 model.

\paragraph{Summary}
Besides the tracers discussed above, we identified two promising molecules that are specific to a core model regardless of the component, once the envelope is discarded from the other components.
We assume the observation is done with a high enough resolution  to take into account the central core, the disk, the pseudodisk, and the outflow only.
One of the promising molecules, HNO,  is specific to the core model MU2$\Theta$0, with an abundance in its central core and its outflow at least one order of magnitude higher than in any other core models.
The other promising molecule, H$_2$CN, is specific to the core model MU10$\Theta$45, with an abundance in its central core which is higher by an order of magnitude or more than in other core models.
The abundance of these two molecules is due to a complex time-dependent gas-phase competition between destruction by H, and formation through recombination of cations and electrons.

\begin{sidewaystable}
\centering
\begin{center}
\caption{Chemical tracers of components}
\label{tab:chem_trac_components}
\begin{tabular}{l|lllll|lllll|llll|llll}
\tableline
\tableline
	&	\multicolumn{5}{c}{MU10$\Theta$0}	&									\multicolumn{5}{c}{MU10$\Theta$45}	&									\multicolumn{4}{c}{MU2$\Theta$0}	&									\multicolumn{4}{c}{MU200$\Theta$0}	\\								
	&	Core	&	Out.	&	Disk	&	Pdisk	&	Env.	&	Core	&	Out.	&	Disk	&	Pdisk	&	Env.	&	Core	&	Out.	&			Pdisk	&	Env.	&	Core	&			Disk	&	Pdisk	&	Env.	\\
\hline																																									
CO	&	{\color{red}{6(-05)}}*	&	5(-10)	&	{\color{blue}{2(-10)}}	&	9(-10)	&	1(-07)	&	{\color{red}{6(-05)}}*	&	{\color{blue}{2(-09)}}	&	3(-08)	&	{\color{blue}{2(-09)}}	&	2(-07)	&	{\color{red}{5(-05)}}*	&	6(-07)			&	{\color{blue}{1(-09)}}	&	2(-07)	&	{\color{red}{5(-05)}}*	&			3(-07)	&	{\color{blue}{4(-09)}}	&	1(-07)	\\
CS	&	{\color{red}{9(-11)}}*	&	7(-16)	&	{\color{blue}{1(-17)}}	&	9(-16)	&	1(-11)	&	{\color{red}{2(-10)}}*	&	5(-15)	&	{\color{blue}{3(-15)}}	&	5(-15)	&	2(-11)	&	{\color{red}{4(-11)}}	&	8(-14)			&	{\color{blue}{1(-15)}}	&	1(-11)	&	{\color{red}{5(-10)}}*	&			2(-13)	&	{\color{blue}{3(-15)}}	&	1(-11)	\\
CN	&	2(-17)	&	5(-16)	&	{\color{blue}{5(-18)}}	&	8(-16)	&	{\color{red}{4(-10)}}*	&	{\color{blue}{7(-18)}}	&	1(-14)	&	3(-15)	&	1(-14)	&	{\color{red}{6(-10)}}*	&	{\color{blue}{2(-17)}}	&	5(-15)			&	2(-15)	&	{\color{red}{4(-10)}}*	&	{\color{blue}{4(-17)}}	&			2(-16)	&	1(-16)	&	{\color{red}{3(-10)}}*	\\
HCO$^+$	&	{\color{blue}{3(-19)}}	&	3(-13)	&	2(-14)	&	4(-13)	&	{\color{red}{6(-11)}}*	&	{\color{blue}{1(-19)}}	&	1(-12)	&	1(-13)	&	8(-13)	&	{\color{red}{7(-11)}}*	&	{\color{blue}{2(-19)}}	&	1(-14)			&	4(-13)	&	{\color{red}{6(-11)}}*	&	{\color{blue}{9(-19)}}	&			5(-16)	&	2(-14)	&	{\color{red}{5(-11)}}*	\\
N2H$^+$	&	{\color{blue}{4(-20)}}	&	2(-12)	&	1(-13)	&	3(-12)	&	{\color{red}{4(-10)}}*	&	{\color{blue}{1(-20)}}	&	8(-12)	&	7(-14)	&	7(-12)	&	{\color{red}{4(-10)}}*	&	{\color{blue}{3(-20)}}	&	6(-15)			&	5(-12)	&	{\color{red}{4(-10)}}*	&	{\color{blue}{9(-20)}}	&			2(-16)	&	5(-14)	&	{\color{red}{3(-10)}}*	\\
HCN	&	{\color{red}{4(-06)}}*	&	7(-16)	&	{\color{blue}{3(-16)}}	&	2(-15)	&	2(-10)	&	{\color{red}{4(-06)}}*	&	{\color{blue}{1(-14)}}	&	4(-13)	&	{\color{blue}{1(-14)}}	&	3(-10)	&	{\color{red}{4(-06)}}*	&	2(-10)			&	{\color{blue}{3(-15)}}	&	2(-10)	&	{\color{red}{1(-06)}}*	&			1(-10)	&	{\color{blue}{3(-14)}}	&	2(-10)	\\
HNC	&	{\color{red}{1(-08)}}*	&	3(-16)	&	{\color{blue}{1(-17)}}	&	6(-16)	&	2(-10)	&	{\color{red}{2(-08)}}*	&	8(-15)	&	1(-14)	&	{\color{blue}{7(-15)}}	&	3(-10)	&	{\color{red}{3(-08)}}*	&	1(-11)			&	{\color{blue}{1(-15)}}	&	2(-10)	&	{\color{red}{4(-09)}}*	&			2(-13)	&	{\color{blue}{2(-15)}}	&	2(-10)	\\
HNO	&	6(-13)	&	2(-12)	&	{\color{blue}{1(-13)}}	&	2(-12)	&	{\color{red}{3(-10)}}*	&	{\color{blue}{1(-18)}}	&	9(-12)	&	4(-12)	&	7(-12)	&	{\color{red}{4(-10)}}*	&	1(-10)	&	1(-10)			&	{\color{blue}{3(-12)}}	&	{\color{red}{4(-10)}}	&	{\color{blue}{4(-15)}}	&			4(-12)	&	2(-12)	&	{\color{red}{3(-10)}}*	\\
C2H	&	{\color{blue}{2(-17)}}	&	2(-15)	&	{\color{blue}{2(-17)}}	&	3(-15)	&	{\color{red}{2(-11)}}*	&	{\color{blue}{2(-18)}}	&	2(-14)	&	8(-16)	&	1(-14)	&	{\color{red}{3(-11)}}*	&	{\color{blue}{2(-17)}}	&	4(-15)			&	5(-15)	&	{\color{red}{3(-11)}}*	&	{\color{blue}{2(-17)}}	&			2(-16)	&	5(-16)	&	{\color{red}{2(-11)}}*	\\
H2O	&	{\color{red}{6(-05)}}*	&	4(-14)	&	{\color{blue}{2(-16)}}	&	4(-14)	&	3(-10)	&	{\color{red}{7(-05)}}*	&	5(-13)	&	{\color{blue}{1(-14)}}	&	3(-13)	&	4(-10)	&	{\color{red}{6(-05)}}*	&	7(-11)			&	{\color{blue}{6(-14)}}	&	3(-10)	&	{\color{red}{2(-06)}}*	&			2(-11)	&	{\color{blue}{2(-13)}}	&	3(-10)	\\
HNCO	&	{\color{red}{6(-08)}}*	&	7(-19)	&	{\color{blue}{3(-20)}}	&	9(-19)	&	3(-14)	&	{\color{red}{6(-08)}}*	&	2(-17)	&	{\color{blue}{1(-17)}}	&	{\color{blue}{1(-17)}}	&	4(-14)	&	{\color{red}{6(-08)}}*	&	9(-14)			&	{\color{blue}{2(-18)}}	&	3(-14)	&	{\color{red}{1(-08)}}*	&			7(-14)	&	{\color{blue}{2(-17)}}	&	3(-14)	\\
H$_2$CN	&	7(-12)	&	2(-16)	&	{\color{blue}{6(-19)}}	&	5(-16)	&	{\color{red}{1(-10)}}*	&	{\color{red}{1(-10)}}	&	9(-15)	&	{\color{blue}{3(-17)}}	&	5(-15)	&	{\color{red}{1(-10)}}	&	2(-12)	&	6(-15)			&	{\color{blue}{1(-15)}}	&	{\color{red}{1(-10)}}*	&	3(-12)	&			4(-16)	&	{\color{blue}{1(-16)}}	&	{\color{red}{7(-11)}}*	\\
H$_2$CO	&	{\color{red}{2(-05)}}*	&	2(-12)	&	{\color{blue}{1(-13)}}	&	2(-12)	&	2(-10)	&	{\color{red}{2(-05)}}*	&	9(-12)	&	{\color{blue}{6(-12)}}	&	{\color{blue}{6(-12)}}	&	3(-10)	&	{\color{red}{2(-05)}}*	&	7(-10)			&	{\color{blue}{3(-12)}}	&	3(-10)	&	{\color{red}{6(-06)}}*	&			3(-09)	&	{\color{blue}{2(-11)}}	&	2(-10)	\\
NH$_3$	&	{\color{red}{3(-05)}}*	&	2(-12)	&	{\color{blue}{5(-14)}}	&	4(-12)	&	6(-09)	&	{\color{red}{3(-05)}}*	&	2(-11)	&	{\color{blue}{9(-13)}}	&	1(-11)	&	8(-09)	&	{\color{red}{3(-05)}}*	&	1(-09)			&	{\color{blue}{6(-12)}}	&	6(-09)	&	{\color{red}{2(-06)}}*	&			2(-10)	&	{\color{blue}{7(-12)}}	&	5(-09)	\\
HC$_3$N	&	{\color{red}{2(-09)}}*	&	8(-17)	&	{\color{blue}{4(-18)}}	&	1(-16)	&	4(-13)	&	{\color{red}{2(-09)}}*	&	2(-16)	&	{\color{blue}{1(-16)}}	&	2(-16)	&	7(-13)	&	{\color{red}{1(-09)}}*	&	8(-14)			&	{\color{blue}{1(-16)}}	&	5(-13)	&	{\color{red}{3(-10)}}*	&			5(-14)	&	{\color{blue}{5(-16)}}	&	4(-13)	\\
C$_3$H$_2$	&	{\color{red}{5(-10)}}*	&	7(-15)	&	{\color{blue}{2(-17)}}	&	8(-15)	&	4(-12)	&	{\color{red}{6(-10)}}*	&	5(-14)	&	{\color{blue}{1(-15)}}	&	3(-14)	&	5(-12)	&	{\color{red}{4(-10)}}*	&	2(-13)			&	{\color{blue}{1(-14)}}	&	5(-12)	&	{\color{red}{2(-10)}}*	&			3(-13)	&	{\color{blue}{1(-14)}}	&	4(-12)	\\
CH$_3$OH	&	{\color{red}{9(-06)}}*	&	3(-13)	&	{\color{blue}{6(-15)}}	&	3(-13)	&	3(-11)	&	{\color{red}{9(-06)}}*	&	1(-12)	&	{\color{blue}{1(-13)}}	&	7(-13)	&	4(-11)	&	{\color{red}{9(-06)}}*	&	7(-11)			&	{\color{blue}{4(-13)}}	&	3(-11)	&	{\color{red}{5(-07)}}*	&			6(-11)	&	{\color{blue}{1(-12)}}	&	3(-11)	\\
\tableline
\end{tabular}
\tablecomments{Core, Pdisk, and Env. refer to central core, pseudodisk, and envelope, respectively. $a(-b)$ means $a\times 10^{-b}$ and displays the mean abundances relative to total hydrogen for a selection of species, at the final time of simulation, of all components from each model.
For each model, the highest and the lowest abundance of a given species are respectively in red and blue.
Abundances with a higher value by a factor 10 or more in one unique component for a given model, compared to all other components of this model, are marked by a * symbol.
These marked species are particularly promising to trace the component.}
\end{center}
\end{sidewaystable}

\begin{sidewaystable}
\begin{center}
\caption{Chemical tracers of models}
\label{tab:chem_trac_models}
\begin{tabular}{l|llll|lll|lll|llll}
\tableline
\tableline
	&	\multicolumn{4}{c}{Core}							&	\multicolumn{3}{c}{Outflow}							&	\multicolumn{3}{c}{Disk}					&	\multicolumn{3}{c}{Pseudodisk}							\\
	&	MU10$\Theta$0	&	MU10$\Theta$45	&	MU2$\Theta$0	&	MU200$\Theta$0	&	MU10$\Theta$0	&	MU10$\Theta$45	&	MU2$\Theta$0			&	MU10$\Theta$0	&	MU10$\Theta$45	&	MU200$\Theta$0	&	MU10$\Theta$0	&	MU10$\Theta$45	&	MU2$\Theta$0	&	MU200$\Theta$0	\\
\hline																															
CO	&		&		&		&		&	{\color{blue}{5(-10)}}	&	2(-09)	&	{\color{red}{6(-07)}}*			&	{\color{blue}{2(-10)}}	&	3(-08)	&	{\color{red}{3(-07)}}*	&	{\color{blue}{9(-10)}}	&	2(-09)	&	1(-09)	&	{\color{red}{4(-09)}}	\\
CS	&	9(-11)	&	2(-10)	&	{\color{blue}{4(-11)}}	&	{\color{red}{5(-10)}}	&		&		&				&		&		&		&		&		&		&		\\
CN	&		&		&		&		&		&		&				&		&		&		&		&		&		&		\\
HCO$^+$	&		&		&		&		&	3(-13)	&	{\color{red}{1(-12)}}	&	{\color{blue}{1(-14)}}			&		&		&		&		&		&		&		\\
N$_2$H$^+$	&		&		&		&		&	2(-12)	&	{\color{red}{8(-12)}}	&	{\color{blue}{6(-15)}}			&		&		&		&	3(-12)	&	{\color{red}{7(-12)}}	&	5(-12)	&	{\color{blue}{5(-14)}}	\\
HCN	&		&		&		&		&	{\color{blue}{7(-16)}}	&	1(-14)	&	{\color{red}{2(-10)}}*			&	{\color{blue}{3(-16)}}	&	4(-13)	&	{\color{red}{1(-10)}}*	&		&		&		&		\\
HNC	&	1(-08)	&	2(-08)	&	{\color{red}{3(-08)}}	&	{\color{blue}{4(-09)}}	&	{\color{blue}{3(-16)}}	&	8(-15)	&	{\color{red}{1(-11)}}*			&		&		&		&		&		&		&		\\
HNO	&	6(-13)	&	{\color{blue}{1(-18)}}	&	{\color{red}{1(-10)}}*	&	4(-15)	&	{\color{blue}{2(-12)}}	&	9(-12)	&	{\color{red}{1(-10)}}*			&	{\color{blue}{1(-13)}}	&	{\color{red}{4(-12)}}	&	{\color{red}{4(-12)}}	&		&		&		&		\\
C$_2$H	&		&		&		&		&		&		&				&		&		&		&		&		&		&		\\
H$_2$O	&	6(-05)	&	{\color{red}{7(-05)}}	&	6(-05)	&	{\color{blue}{2(-06)}}	&	{\color{blue}{4(-14)}}	&	5(-13)	&	{\color{red}{7(-11)}}*			&	{\color{blue}{2(-16)}}	&	1(-14)	&	{\color{red}{2(-11)}}*	&		&		&		&		\\
HNCO	&		&		&		&		&		&		&				&		&		&		&		&		&		&		\\
H$_2$CN	&	7(-12)	&	{\color{red}{1(-10)}}*	&	{\color{blue}{2(-12)}}	&	3(-12)	&		&		&				&		&		&		&		&		&		&		\\
H$_2$CO	&	{\color{red}{2(-05)}}	&	{\color{red}{2(-05)}}	&	{\color{red}{2(-05)}}	&	{\color{blue}{6(-06)}}	&	{\color{blue}{2(-12)}}	&	9(-12)	&	{\color{red}{7(-10)}}*			&	{\color{blue}{1(-13)}}	&	6(-12)	&	{\color{red}{3(-09)}}*	&	{\color{blue}{2(-12)}}	&	6(-12)	&	3(-12)	&	{\color{red}{2(-11)}}*	\\
NH$_3$	&	{\color{red}{3(-05)}}	&	{\color{red}{3(-05)}}	&	{\color{red}{3(-05)}}	&	{\color{blue}{2(-06)}}	&	{\color{blue}{2(-12)}}	&	2(-11)	&	{\color{red}{1(-09)}}*			&	{\color{blue}{5(-14)}}	&	9(-13)	&	{\color{red}{2(-10)}}*	&		&		&		&		\\
HC$_3$N	&	{\color{red}{2(-09)}}	&	{\color{red}{2(-09)}}	&	1(-09)	&	{\color{blue}{3(-10)}}	&		&		&				&		&		&		&		&		&		&		\\
C$_3$H$_2$	&		&		&		&		&		&		&				&		&		&		&		&		&		&		\\
CH$_3$OH	&	{\color{red}{9(-06)}}	&	{\color{red}{9(-06)}}	&	{\color{red}{9(-06)}}	&	{\color{blue}{5(-07)}}	&	{\color{blue}{3(-13)}}\	&	1(-12)	&	{\color{red}{7(-11)}}			&	{\color{blue}{6(-15)}}	&	1(-13)	&	{\color{red}{6(-11)}}*	&		&		&		&		\\
\tableline
\end{tabular}
\tablecomments{Same as Table~\ref{tab:chem_trac_components} but for model sensitivity.
Among these gas phase species, only the ones with an abundance $\geq 10^{-12}$ in one or more models for a given component and with a significant sensitivity to the model (minimum of about a factor 10) are displayed to facilitate the reading.
For each component, the highest and the lowest abundance of a given species are respectively in red and blue.
Abundances with a higher value by a factor 10 or more in one unique model for a given component, compared to all other models for this component, are marked by the * symbol.
These marked species are particularly promising to trace the model.
The envelope does not show any tracer allowing a differentiation between models.}
\end{center}
\end{sidewaystable}

\section{Comparison with recent observations}
\label{sec:obs}

\cite{gerin_nascent_2015} observed H$_2$CO and CH$_3$OH in the outflow of the two first Larson core candidates Barnard 1b-N and 1b-S, which have a high magnetization level ($2<\mu<7$) and an inclined magnetic field.
They derived a methanol abundance relative to total hydrogen nuclei of $5\times 10^{-9}$ and $3.5\times 10^{-8}$ respectively.
Our results show the abundance of methanol is higher for a higher magnetization level (on average by a factor of 200), and is also higher if the magnetic field is inclined (by a factor of 4), which may explain the high observed abundance.
We obtain  mean abundances of $3\times 10^{-13}$, $1\times 10^{-12}$, and $7\times 10^{-11}$ and  maximum abundances of $6\times 10^{-12}$, $1\times 10^{-10}$, and $9\times 10^{-6}$ for MU10$\Theta$0, MU10$\Theta$45, and MU2$\Theta$0 respectively.
Note that a probably more adequate model for this specific source would contain $2<\mu<7$ and $\Theta \neq 0$.

{\cite{huang_probing_2013} and \cite{hirano_two_2014} also observed these two sources.
They derived the same N$_2$H$^+$ abundance and a similar HCO$^+$ abundance to what we obtained in the collapsing envelope.
They also suggest a depletion of CO in the dense gas around the two sources, as we obtain in the envelope.
}

{\cite{tsitali_dynamical_2013} observed the first Larson core candidate Chamaeleon-MMS1.
They derived abundance profiles for CS, HCO$^+$, and CO that show a decrease from the outer radius ($\sim 10^4$~AU) to the inner radius ($\sim 20$~AU).
The abundances vary from $10^{-9}$, $10^{-8}$, and $10^{-5}$ in the outer radius to $10^{-15}$, $10^{-11}$, and $10^{-7}$ in the inner radius, for CS, HCO$^+$, and CO, respectively.
Our results show the same values in the outer radius, and the same trend towards the inner radius.

\cite{jaber_census_2014} derived abundances of several COMs from observation of the cold envelope of IRAS16293, a solar-type Class~0 protostar.
Although our simulation stops before the second Larson core, the cold envelope may not evolve drastically so it is worthy to compare with these recent observations.
The observed COMs with their derived abundances are H$_2$CCO ($\sim 10^{-11}$ relative to the total proton density), CH$_3$CHO ($\sim 10^{-9}$), NH$_2$CHO ($\sim 10^{-12}$ to $\sim 10^{-10}$), CH$_3$OCH$_3$ ($\sim 10^{-10}$ to $\sim 10^{-8}$), and HCOOCH$_3$ ($\sim 10^{-11}$ to $\sim 10^{-9}$).
A range in the results may exist because of the radial dependence of the abundance, from $\sim$100~AU to $\sim$10,000~AU.
Our models reproduce the observation of H$_2$CCO and NH$_2$CHO but not for the other molecules because we do not yet include the new proposed gas-phase or surface routes to form them (see the discussion about COMs in Section~\ref{subsubsec:distri_molecules}).

\cite{kwon_kinematics_2015} observed N$_2$H$^+$ lines towards the young Class~0 protostellar system L1157.
They identified the outer collapsing envelope with a double-peaked feature and observed a central hole in the N$_2$H$^+$ flux, as suggested by our computed abundances of N$_2$H$^+$ in the central core and the collapsing envelope, as shown in Table~\ref{tab:chem_trac_components}.

\section{Summary and conclusion}
\label{sec:conclusion}

We have followed the physical and chemical evolution of several collapsing prestellar dense cores, from the molecular cloud stage to the formation of the first hydrostatic core, also called the first Larson core.
To do so, we coupled the full gas-grain chemical code {\ttfamily{NAUTILUS}} with the radiative magneto hydrodynamical code {\ttfamily{RAMSES}}.
{\ttfamily{RAMSES}} provides a detailed 3D time-dependent physical structure, while {\ttfamily{NAUTILUS}} allows us to follow the full gas-grain chemistry for the majority of the existing interstellar molecules.
The combination of both is thus a powerful tool to study astrophysical objects, such as dense cores in this present study.
Both chemistry and physics evolve as a function of time\footnote{{Note that in our simulation, feedback from the chemistry on the dynamics is not possible.}}, and we emphasize the fact that the chemistry reaches neither equilibrium nor steady state, which makes this time-dependent coupling an important point of our work.
The chemical timescale is often not negligible compared with the dynamical timescale and vice-versa, which indicates the importance of modeling both chemical and dynamical evolution in parallel.

We used different assumptions concerning the magnetic field of these cores, namely on its intensity and inclination.
We realized in total four different simulations, from a quasi pure hydrodynamical simulation to a highly magnetized core simulation.
Each collapsing core can be divided in up to five different components: a central core (the first Larson core), an outflow, a disk, a pseudodisk, and an envelope.
After a separation of these components using criteria on the velocity field, the kinetic energy and the thermal energy, we have studied the physical and chemical characteristics of each component, for each model, and highlighted the noticeable differences.

We first described the size of the components at the final step of our simulations, and their density and temperature conditions, and then explored their prior history, during the collapse.
Besides their final state, each component often has a specific and unique physical history, which can even be different from the same component of a different model.

Then, we described the distribution of the molecules, namely the principal reservoirs of carbon, nitrogen, and oxygen, and the charged species and complex organic molecules as well.
We also took a closer look at the chemical evolution along a few representative trajectories.
Even though the duration of the collapse is short, a few times $10^4$~yr, the chemistry may have time to notably change the abundances of some molecules, such as methyl formate, on the grain surface.
This is mainly due to the high density condition, which greatly enhances the total rates of the chemical reactions, and the local temperature that may possess an adequate range of values in order to allow a fast diffusion on the grain surface without desorption.
We also find that for two tracer particles that share the same initial and the same final physical conditions, their final chemical content may differ by a factor of a few  due to a different past dynamical history along their respective trajectories.
This distinction may limit the possibility of deriving a precise dynamical history from observed abundances.

Since the first Larson core has not been detected with firm confirmation, we tried to identify some chemical signatures that could help its identification.
We performed a general study to discriminate those species with abundances most sensitive to the considered component and model, and identified those that are the best tracer candidates.
While the collapsing envelopes present very similar chemical compositions from one model to the other, this is not the case for the other components.
{We therefore recommend to use a high enough resolution (i.e. through interferometry) to observe the inner components within the collapsing envelope, and not focus on the envelope itself.}
Some species have an enhanced abundance in a given component, while they are depleted in other components.
Besides, the different models may present significant chemical variation for a given component as well.
{Some results for gas-phase species, itemized below, are particularly important:
\begin{enumerate}
\item CN, HCO$^+$, N$_2$H$^+$, HNO, C$_2$H, and H$_2$CN trace the envelope, while the other species discussed in section~\ref{subsec:tracers} trace the central core;
\item NH$_3$, N$_2$H$^+$, and HCO$^+$  are useful in distinguishing the disk from the pseudodisk;
\item a majority of the species discussed in section~\ref{subsec:tracers} -- CS, CN, HCO$^+$, N$_2$H$^+$, HNC, C$_2$H, H$_2$CN, and C$_3$H$_2$ -- are useful in distinguishing the envelope from the outflow or the pseudodisk; 
\item the chemistry of the envelope cannot help to make a distinction between different models of the pres-tellar cores;
\item the abundances of species are generally higher in the outflow of a highly magnetized core, which implies that chemistry may be a clue to know more about the intensity of the magnetic field;
\item HNO and H$_2$CN are unusually abundant in the central core of the MU10$\Theta$0 model and the MU10$\Theta$45 model, respectively, which means they may give information on the strength and the inclination of the magnetic field;
\item CO and HNO allow a good distinction between the disks of the MU10$\Theta$0 and the MU10$\Theta$45 models.
\end{enumerate}}

{Chemistry is thus a promising tool to infer characteristics of first Larson cores and their assorted components.
To have a better quantitative estimation of the observability of the tracer chemical species, the generation of synthetics maps of molecular emission will be relevant, and will be the next step towards a more complete characterization of the collapsing pre-stellar object at the time of the first Larson core.}

\acknowledgments

The authors thank the anonymous referee for  a careful reading of the manuscript and for suggested modifications, which allowed us to improve the initial version of the paper.
This research was partially funded by the program PCMI from CNRS/INSU.
UH was funded by a grant from the French ''R\'egion Aquitaine'' when part of this work was conducted.
UH and EH acknowledge support from the National Science Foundation through a grant to EH.
The research of BC is supported by the ANR Retour Postdoc program and by the CNES.
BC acknowledges postdoctoral fellowship support from the Max-Planck-Institut f\H{u}r Astronomie where part of this work was conducted.
{The research of VW and FH is funded by an ERC Starting Grant (3DICE, grant agreement 336474).}
The {\ttfamily{RAMSES}} calculations have been performed  at CEA on the DAPHPC cluster and the {\ttfamily{NAUTILUS}} calculations were performed using JADE cluster resources from GENCI-CINES (Grand Equipement National de Calcul Intensif - Centre Informatique National de l'Enseignement Sup\'erieur).
Some kinetic data we used were downloaded from the online database KIDA (KInetic Database for Astrochemistry, \url{http://kida.obs.u-bordeaux1.fr}, \cite{wakelam_kinetic_2012}).
GDL \citep{coulais_status_2010,coulais_space_2012} and VisIt \citep{childs_visit:_2012} have been used to process data.

\bibliographystyle{aa} 
\bibliography{ma_biblio} 

\appendix

\section{A. Physical history of particles: temperature and density distributions of particles}
\label{appendix:histogram}
Figures~\ref{fig:appendix_histogram_mu10theta0} to \ref{fig:appendix_histogram_mu2theta0} display normalized distributions of the temperature and the density of the particles that belong at the end of the simulations to the central core, the outflow, the disk, the pseudodisk, and the envelope, for all the models.

\begin{figure}[h]
\centering
\includegraphics[width=1.0\linewidth]{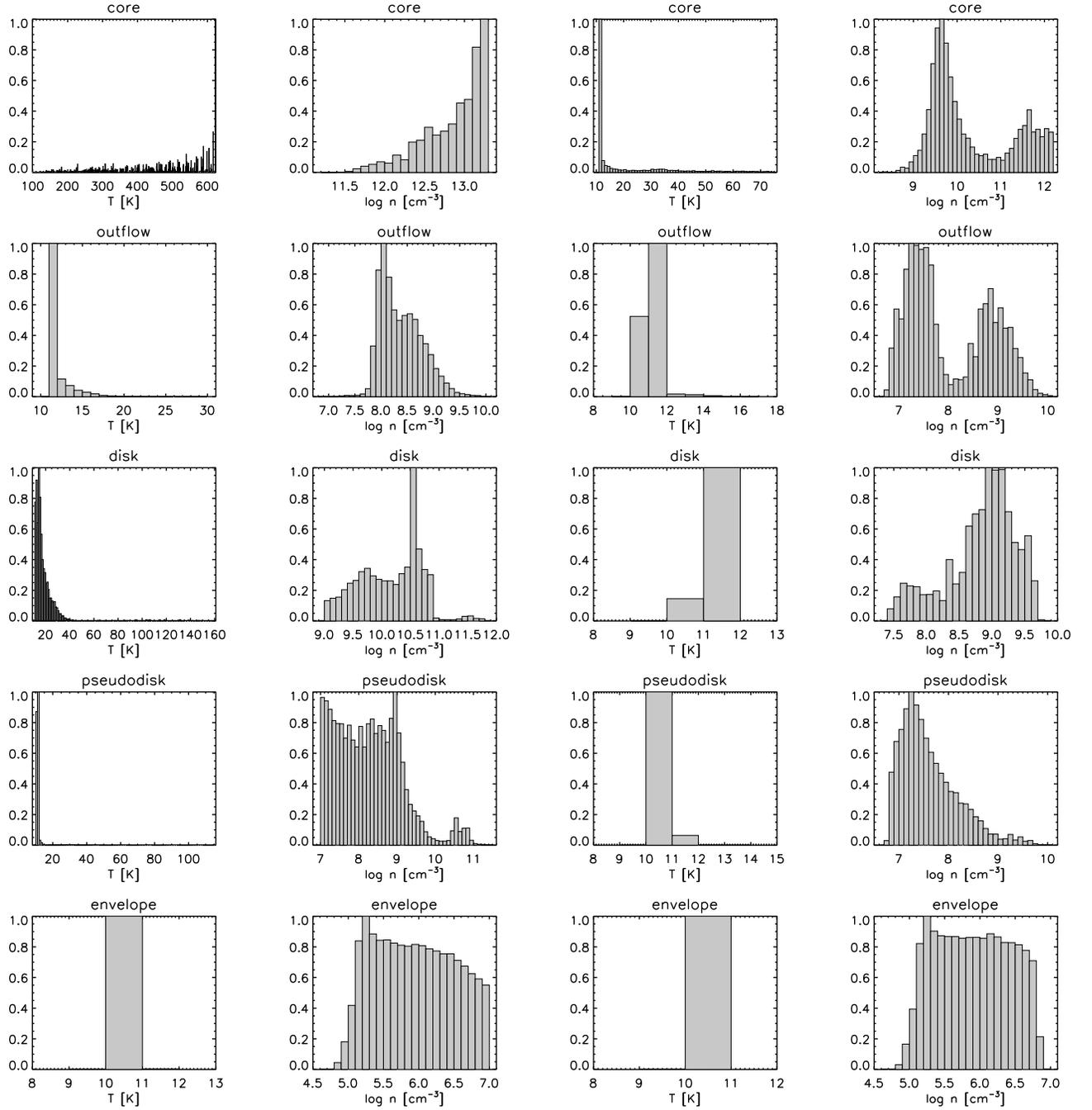}
\caption{Normalized distributions of temperature and density of particles that belong at the end of the simulations to the central core, the outflow, the disk, the pseudodisk, and the envelope, for the MU10$\Theta$0 model.
The two left columns and two right columns correspond respectively to the final time, and to $t_0$ (see text in section~\ref{subsubsec:phys_history_matter}).}
\label{fig:appendix_histogram_mu10theta0}
\end{figure}

\begin{figure}[h]
\centering
\includegraphics[width=1.0\linewidth]{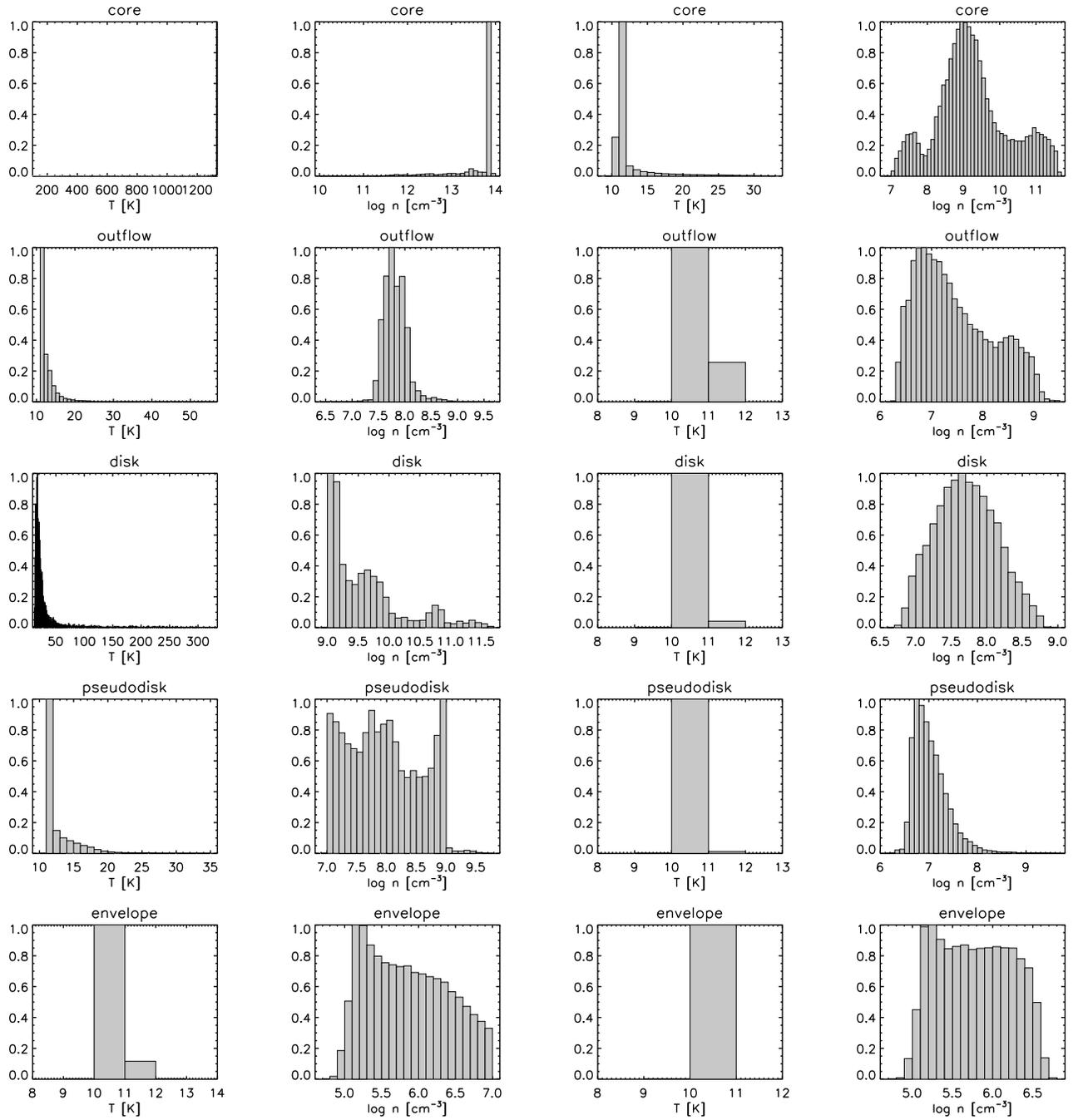}
\caption{Same as Figure~\ref{fig:appendix_histogram_mu10theta0} for the MU10$\Theta$45 model.}
\label{fig:appendix_histogram_mu10theta45}
\end{figure}

\begin{figure}[h]
\centering
\includegraphics[width=1.0\linewidth]{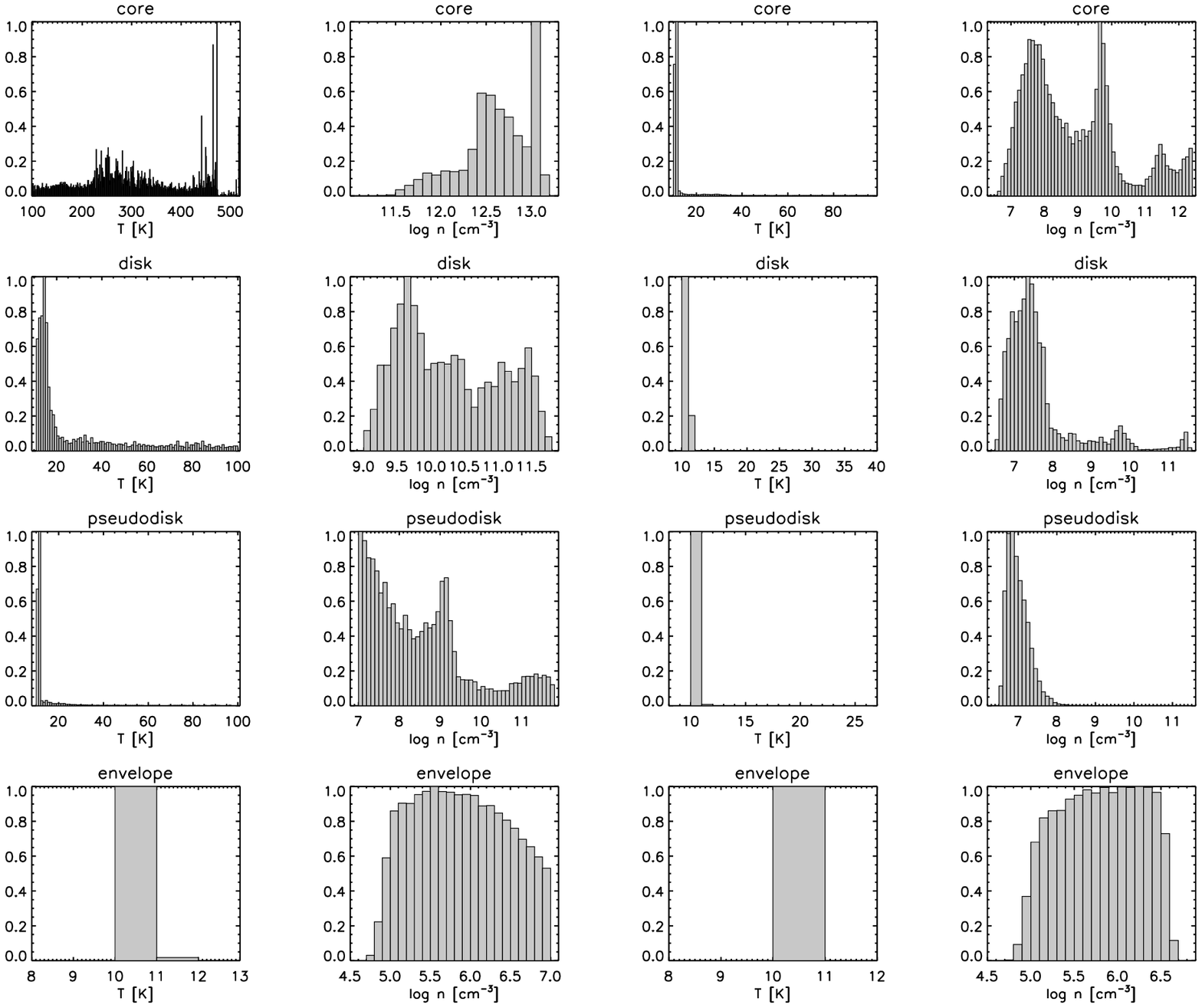}
\caption{Same as Figure~\ref{fig:appendix_histogram_mu10theta0} for the MU200$\Theta$0 model.}
\label{fig:appendix_histogram_mu200theta0}
\end{figure}

\begin{figure}[h]
\centering
\includegraphics[width=1.0\linewidth]{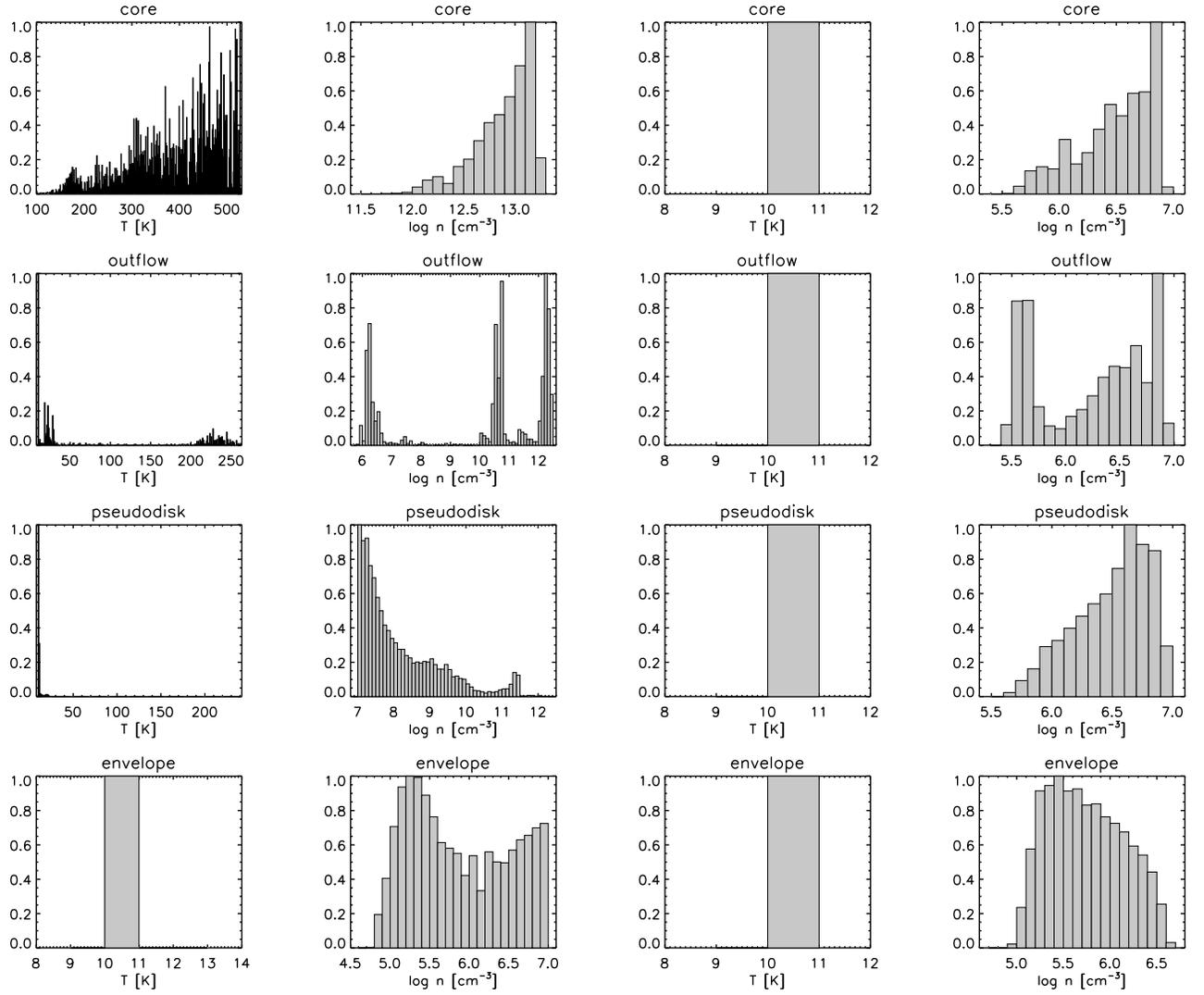}
\caption{Same as Figure~\ref{fig:appendix_histogram_mu10theta0} for the MU2$\Theta$0 model.}
\label{fig:appendix_histogram_mu2theta0}
\end{figure}

\clearpage
\section{B. Adsorption and thermal desorption timescales}
\label{appendix:timescales}

Adsorption and thermal desorption are the major processes that affect the evolution of the chemical material in our simulations, due to the high density and the short collapsing time.
The efficiency of these two processes depend both on physical parameters, such as the temperature, the density, and the gas to dust ratio, and on chemical parameters, such as the desorption energy and the mass of the species.
The characteristic timescales of these processes reflect their efficiency, and the difference between them tell us if chemical species are mainly present in the gas phase or on the grain surface.
Thus, we present a detailed calculation of adsorption and thermal desorption timescales.

\subsection{B.1. Adsorption timescale}
\label{appendix:adsorption_timescale}

The adsorption of a molecule $X$ on a grain surface can be represented by the reaction
\begin{equation}
grain + X \longrightarrow JX,
\label{eq:ads_reaction}
\end{equation}
where $X$ and $JX$ represent the molecule respectively in the gas phase, and on the grain surface.
The evolution of the gas phase density of $X$, $n_X(t)$, is then given by
\begin{equation}
\frac{d}{dt}n_X(t) = -k_\mathrm{ads} n_X(t) n_\mathrm{d},
\label{eq:ads_reaction_evolution}
\end{equation}
where $k_{ads}$ is the adsorption rate constant related to \ref{eq:ads_reaction}, $n_\mathrm{d}$ is the grain density, and $t$ is the time.
The adsorption characteristic timescale $\tau_\mathrm{ads}$ can be defined as
\begin{equation}
\tau_\mathrm{ads}=\left|\frac{n_X(t)}{\frac{d}{dt}n_X(t)}\right|.
\label{eq:timescale}
\end{equation}
Using equations~\ref{eq:ads_reaction_evolution} and \ref{eq:timescale}, we find that
\begin{equation}
\tau_\mathrm{ads}=\frac{1}{k_\mathrm{ads}n_\mathrm{d}}.
\end{equation}
The adsorption rate constant is a function of the grain cross section $\sigma_\mathrm{d}$, the thermal velocity of the gas $<v(i)>$, and the adsorption probability $P_\mathrm{ads}$ 
\begin{equation}
k_\mathrm{ads} = \sigma_\mathrm{d} <v(i)> P_\mathrm{ads}.
\end{equation}
If we assume that the gas velocity has a Maxwell-Boltzmann distribution, the rate is written as
\begin{equation}
k_\mathrm{ads} = \sigma_\mathrm{d} \sqrt{\frac{8k_\mathrm{B}T}{\pi m(X)}} P_\mathrm{ads},
\end{equation}
where $T$ is the gas temperature and $m(X)$ is the mass of the species $X$.

Grain density is linked to the gas density $n_{\mathrm H_2}$, the mean molecular mass $\mu_m$, the proton mass $m_\mathrm{p}$, the gas to dust ratio by mass $R_\mathrm{gd}$, and the grain mass $m_\mathrm{d}$ by
\begin{equation}
n_\mathrm{d} = \frac{n_{\mathrm H_2} \mu_m m_\mathrm{p}}{R_\mathrm{gd} m_\mathrm{d}}.
\end{equation}

The adsorption timescale is then a function of the temperature and the density of the matter (that themselves depend on the time $t$), and the considered species (through its mass)
\begin{equation}
\boxed{ \tau_\mathrm{ads} = \sqrt{\frac{\pi}{8 k_B}} \frac{R_{gd} m_d}{\sigma_d P_{ads} \mu_m m_p} \frac{1}{n_{\mathrm H_2}(t)} \sqrt{\frac{m(X)}{T(t)}} }.
\end{equation}

The adsorption probability is defined as the probability for a gas phase species to thermally equilibrate with the grain.
In our simulations, we use an adsorption probability equal to 1 \citep{dhendecourt_time_1985}, even though it is dependent on some parameters such as the gas and grain temperatures \citep[see for example][]{fillion_gas-surface_2011}, or the composition and the structure of the grain surface \citep[see for example][]{buch_sticking_1991}.
Experimental studies show that adsorption probability of heavy molecules (O$_2$, N$_2$, CO, CH$_4$, and H$_2$O) exceed 0.9 \citep{kimmel_control_2001,bisschop_desorption_2006,acharyya_desorption_2007}.
For light species, \cite{hollenbach_surface_1970}, \cite{buch_sticking_1991}, \cite{masuda_sticking_1998}, and \cite{al-halabi_hydrogen_2007} show that the adsorption probability is between 0.8 and 1, even though it depends on the grain surface temperature, and the incidence energy of the molecule to the surface.

\subsection{B.2. Thermal desorption timescale}
\label{appendix:desorption_timescale}

The desorption of a molecule $JX$ from the grain surface can be represented by the reaction
\begin{equation}
JX \longrightarrow grain + X.
\label{eq:des_reaction}
\end{equation}
The evolution of the grain surface density of the molecule, $n_{JX}(t)$, is then given by
\begin{equation}
\frac{d}{dt}n_{JX}(t) = -k_\mathrm{des} n_{JX}(t)
\label{eq:des_reaction_evolution}
\end{equation}
where $k_\mathrm{des}$ is the desorption rate constant related to \ref{eq:des_reaction}.
Using a similar definition as \ref{eq:timescale}, the desorption characteristic timescale $\tau_\mathrm{des}$ is
\begin{equation}
\tau_\mathrm{des} = \frac{1}{k_\mathrm{des}}.
\end{equation}
The desorption rate constant is a function of the desorption energy $E_\mathrm{D}(JX)$\footnote{
We assume that chemical species on grain are physisorbed on water ice, the main component of the grain mantle (see for example \cite{hasegawa_new_1993} and \cite{cuppen_simulation_2007} for more details on desorption energy).
}, the grain temperature $T_d$, and the characteristic adsorbate vibrational frequency $\nu(JX)$ 
\begin{equation}
k_\mathrm{des} = \nu(JX) \exp\left(\frac{- E_\mathrm{D}(JX)}{T_\mathrm{d}}\right),
\end{equation}
where $\nu(JX)$ is given by
\begin{equation}
\nu(JX) = \sqrt{\frac{2 n_s E_\mathrm{D}(JX)}{\pi^2 m(JX)}},
\label{eq:char_ads_vib_freq}
\end{equation}
$n_\mathrm{s}$ is the grain site surface density.

The desorption timescale is then a function of the chemical species (through its mass, and its desorption energy), and the temperature of the matter :
\begin{equation}
\boxed{ \tau_\mathrm{des} = \sqrt{\frac{\pi^2}{2n_s}} \sqrt{\frac{m(JX)}{E_\mathrm{D}(JX)}} \exp\left(\frac{E_\mathrm{D}(JX)}{T(t)}\right) }.
\end{equation}

\clearpage
\section{C. Chemical steady-state versus chemical equilibrium}
\label{appendix:ss_vs_eq}

Chemical steady-state and chemical equilibrium are two different concepts.
Consider the left-to-right reaction
\begin{equation}
\rm A + B \longrightarrow C + D
\end{equation}
with a rate coefficient $k_f$, where$f$ stands for "forward reaction".
For the backward (right-to-left) reaction:
\begin{equation}
\rm C + D \longrightarrow A + B.
\end{equation}
we label the rate coefficient $k_b$.
Now consider the forward and backward reactions to be part of a larger system with other reactions.  The concept of steady-state refers to a situation in which
the concentrations of all reactants and the products will be constant, which means that the time derivatives of the concentrations of [A], [B], [C], and [D] are zero:
\begin{equation}
\label{eq:steady_state}
\frac{d[{\rm A}]}{dt} = \frac{d[{\rm B}]}{dt} = \frac{d[{\rm C}]}{dt} = \frac{d[{\rm D}]}{dt} = 0.
\end{equation}
and the same condition is reached for all other species in the system.  
Such a condition can be obtained if for each molecule in the system, the sum of the rates of assorted reactions to form it is equal to the sum of the rates of assorted reactions to destroy it.   The existence of the steady-state condition does not require that the forward and backward reactions have the same rate.  But chemical equilibrium does require that the forward and backward reactions do have the same overall rate for each reaction in the system.  If we consider the reactions between A and B and between C and D, chemical equilibrium requires that
\begin{equation}
k_f [{\rm A}] [{\rm B}] = k_b [{\rm C}] [{\rm D}],
\end{equation}
where the equilibrium constant $K$ is defined by the ratio of the rate coefficients, or by the concentrations of the reactants and products:
\begin{equation}
K = \frac{k_{f}}{k_{b}} =    \frac{[{\rm C}] [{\rm D}]}{[{\rm A}] [{\rm B}]}.
\end{equation}
Expressed in thermodynamic terms, the overall system reaches equilibrium when the so-called Gibbs free energy achieves a minimum value.  
Another thermodynamic expression at equilibrium involves only individual reaction pairs such as our four species -- A, B, C, and D:  
\begin{equation}
K = \exp \left( - \frac{\Delta_rG^o}{RT} \right),
\end{equation}
where the equilibrium constant is a function of the difference between the  Gibbs free energy of products and reactants, $\Delta_rG^o$, at their standard states,  and the temperature $T$,
where $R$ is the gas constant.

Chemical equilibrium usually needs very high temperatures -- more than 1,000~K -- to allow all forward and backward reactions to determine the concentrations of reactants and products in terms of equilibrium constants, so that both forward and backward reactions can overcome an activation energy barrier or a simple endothermicity for one of the directions.
For example, it is used to model the gas-phase chemistry of the hot atmosphere of giant planets \citep{burrows_chemical_1999}.  Individual forward-backward pairs of reactions can reach equilibrium at lower temperatures under unusual conditions such as the system
\begin{equation}
\rm H_3^+ + HD \rightleftarrows H_2D^+ + H_2
\end{equation}
as long as the forward and backward reactions dominate the chemical formation and destruction. In this system, the left-to-right reaction is exothermic by only 230 K.  Under typical dense interstellar conditions, however, the species H$_{2}$D$^{+}$ can be destroyed more rapidly by reaction with electrons and CO than by the endothermic backwards reaction.  

\end{document}